\documentclass[traditabstract]{aa}

\usepackage{graphicx}
\usepackage{txfonts}
\usepackage{enumerate}

\usepackage{natbib}
\bibpunct{(}{)}{;}{a}{}{,} 

\begin{document} 

\title{Fitting Isochrones to Open Cluster photometric data}
\subtitle{A new global optimization tool}

\author{H. Monteiro\thanks{E-mail:
    hektor.monteiro@gmail.com}, W. S. Dias and T. C. Caetano }
  \institute{UNIFEI, DFQ - Instituto de Ci\^encias Exatas, Universidade Federal de Itajub\'a, Itajub\'a MG, 
Brazil}

\date{}

\begin{abstract}
  {We present a new technique to fit color-magnitude diagrams of open
    clusters based on the Cross-Entropy global optimization
    algorithm. The method uses theoretical isochrones available in the
    literature and maximizes a weighted likelihood function based on
    distances measured in the color-magnitude space. The weights are
    obtained through a non parametric technique that takes into
    account the star distance to the observed center of the cluster,
    observed magnitude uncertainties, the stellar density profile of
    the cluster among others. The parameters determined simultaneously
    are distance, reddening, age and metallicity. The method takes
    binary fraction into account and uses a Monte-Carlo approach to
    obtain uncertainties on the determined parameters for the cluster
    by running the fitting algorithm many times with a re-sampled data
    set through a bootstrapping procedure. We present results for 9
    well studied open clusters, based on 15 distinct data sets, and
    show that the results are consistent with previous studies. The
    method is shown to be reliable and free of the subjectivity of
    most previous visual isochrone fitting techniques.}
\end{abstract}

\keywords{open clusters and associations: general.} 

\maketitle

\section{Introduction}

Galactic open clusters are a key class of objects used in a wide range
of investigations due to their wide span of ages and distances as well
as the precision to which these parameters can be determined through
their color-magnitude diagrams (CMD). With the publication of the
Hipparcos Catalog (ESA 1997) and its derivatives the Tycho and Tycho2
\citep{ESA1997,Hog2000} as well as individual efforts using modern
ground based instrumentation there has been increased interest in
studies involving open clusters in the Galaxy.

Individually the results obtained from the study of open clusters can
provide important constraints for theoretical models of stellar
formation and evolution. Comparison of observed CMDs to model
isochrones can provide important information on the effect of
overshooting \citet{VandenBerg2004} and chemical abundances
\cite{Meynet1993, Kassis1997}. Open clusters can also be used in the
study of variable stars, especially Cepheids
(\cite{Kang2008,Majaess2009}), in the search for connections between
stellar magnetic fields and their evolution and in the search for
extra-terrestrial planets, among many others.

Our group has focused efforts in the investigation of open clusters
and their use in the understanding of the Galactic spiral structure
(\cite{Dias2005,lepine2008}).  The results are based on the catalog of
open clusters published in \cite{dias2002} witch is now in version
2.10 and, since 2002, can be accessed on line at
http://www.astro.iag.usp.br/$\sim$wilton.

The results presented in our catalog are compiled from data published
from many authors, using different instruments, techniques,
calibrations and criteria which results in a heterogeneous
sample. However, in \cite{PN06}, the authors show that our data has
the same statistical significance as the data they use to define the
standard parameters of the chosen open clusters when considering the
calculated errors.

To derive the fundamental parameters of open clusters the main
sequence ``fitting'' in most works up to today has been done mostly
with subjective visual fitting. This is mainly due to the fact that
the isochrones have no simple parametric form so that a usual least
square technique can be applied. The available isochrones in the
literature are usually in the form of tabulated points for a set of
fundamental parameters such as age and metallicity. The traditional
``fitting'' method utilized is to first determine the reddening by
adjusting a Zero Age Main Sequence (ZAMS) to the observed color-color
diagram (usually $(B-V)$ vs. $(U-B)$ of the cluster and then, keeping
this value fixed, adjusting the distance and age using the observed
CMD and tabulated isochrones. Both of these steps are usually
performed by eye as mentioned, leading to inevitable subjectivity in
the fitting of an isochrone to a given observed CMD.  The subjectivity
in the determination of the reddening is specially problematic since
it affects the subsequent determination of the distance and age of the
cluster. It is also important to point out that in some cases the
reddening is not determined through the use of U filter photometry
which can further compromise the results obtained.

The lack of homogenization of the data and especially the subjectivity
of the methods used to obtain fundamental parameters of clusters from
their CMD indicate the great need for a method that circumvents at
least the subjectivity. The subject of automatic fitting and non
subjective criteria to choose the best fit was addressed in some
papers such the recently published work by \cite{Naylor2006}, where
the authors propose a maximum likelihood method for fitting two
dimensional model distributions to stellar data in a CMD. In this work
the authors also discuss the most important attempts at performing
isochrone fits using different methods and we refer the reader to
their very complete discussion and references therein. The authors
also present the main problems involved in this kind of study, as the
data precision (see also \cite{Narbutis2007}), the importance of non
resolved binaries (see \cite{Schlesinger1975, Fernie1961} and in more
modern observational studies \cite{Montgomery1993, vonHippel1998}).

In this work we present a new technique to fit models to open cluster
photometric data using a weighted likelihood criterion to define the
goodness of fit and a global optimization algorithm known as
Cross-Entropy to find the best fitting isochrone. We have successfully
applied the optimization algorithm in the study of jet precession in
Active Galactic Nuclei (\cite{CMA09}) where we demonstrate the
robustness of the method. In the present work we adapt the method to
study open clusters and show that it can find the best parameters and
eliminate the subjectivity (author analysis dependence) of the main
sequence fitting process by using well defined control parameters and
a weighted likelihood.

In the next section we introduce the optimization technique and how it
was adapted to the problem of ishocrone fitting. In Sec 3. we define
the likelihood function used and in Sec. 4 how the weight function is
obtained. In Sec. 5 we demonstrate the validity of the method by
applying it to synthetic clusters and in Sec. 6 we apply the method to
the data of 9 clusters carefully chosen. In Sec. 7 we discuss the
results and in Sec. 8 we give our final conclusions.

\section[]{Cross entropy global optimization}

\subsection{The optimization method}

The Cross Entropy technique (CE) was first introduced by
\cite{rubi97}, with the objective of estimating probabilities of rare
events in complex stochastic networks, having been modified later by
\cite{rubi99} to deal with continuous multi-extremal and discrete
combinatorial optimization problems. Its theoretical asymptotic
convergence has been demonstrated by \cite{marg04}, while
\cite{kro06} studied the efficiency of the CE method in solving
continuous multi-extremal optimization problems. Some examples of
robustness of the CE method in several situations are listed in
\cite{deb05}. The CE procedure uses concepts of importance sampling,
which is a variance reduction technique, but removing the need for a
priory knowledge of the reference parameters of the parent
distribution. The CE procedure provides a simple adaptive way of
estimating the optimal reference parameters. Basically, the CE method
involves an iterative procedure where in each iteration the following
is done:

\begin{enumerate}[i]
\item Random generation of the initial parameter sample, respecting
  pre-defined criteria;
\item Selection of the best candidates based on some mathematical
  criterion;
\item Random generation of updated parameter samples from the previous
  best candidates to be evaluated in the next iteration;
\item Optimization process repeats steps (ii) and (iii) until a
  pre-specified stopping criterion is fulfilled.
\end{enumerate}

The CE algorithm is based on a population of solutions in a similar
manner to the well known genetic algorithm and in this sense it is
also a type of evolutive algorithm where some fraction of the
population is selected in each iteration based on a given selection
criteria. However, a detailed discussion of evolutive algorithms and
their similarities and comparative performance is beyond the scope of
this work. In the work of \cite{deb05}, many standard benchmark
optimization problems, such as the Travelling Salesman, are studied
using the CE method and its efficiency is also discussed. A great
advantage of the CE method over the genetic algorithm for example is
its simplicity to code. For problems with many free parameters there
is no need to deal with genes and their definitions, crossing,
mutation rates and other details.

Here we have implemented the CE method to find the best
fitting model isochrone to open cluster data as discussed in the
following sections.

\subsection[]{Cross entropy isochrone fitting}

Let us suppose that we wish to study a set of $N_{d}$
observational data in terms of an analytical model characterized by
$N_{p}$ parameters $p_1, p_2, ..., p_{N_{p}}$.

The main goal of the CE continuous multi-extremal optimization method
is to find a set of parameters $p^*_i$ ($i=1, ..., N_{p}$) for
which the model provides the best description of the data
(\cite{rubi99,kro06}). It is performed generating randomly $N$
independent sets of model parameters ${\bf X}=({\bf x}_1,{\bf
  x}_2,...,{\bf x}_N)$, where ${\bf
  x}_i=(p_{1_i},p_{2_i},...,p_{N_{{p}i}})$, and minimizing the
objective function $S({\bf X})$ used to transmit the quality of the
fit during the run process. If the convergence to the exact solution
is achieved then $S\rightarrow 0$, which means ${\bf x}\rightarrow{\bf
  x}^*=(p^*_1,p^*_2,...,p^*_N)$.

In order to find the optimal solution from CE optimization, we start
by defining the parameter range in which the algorithm will search for
the best candidates: $\xi^{min}_i\leq p_i \leq
\xi^{max}_i$. Introducing
$\bar{\xi}_i(0)=(\xi^{min}_i+\xi^{max}_i)/2$ and
$\sigma_i(0)=(\xi^{max}_i-\xi^{min}_i)/2$, we can compute
${\bf X}(0)$ from:

\begin{equation}
  X_{ij}(0)=\bar{\xi}_i(0)+\sigma_i(0) G_{i,j},
\end{equation}
where $G_{i,j}$ is an $N_{p}\times N$ matrix with random numbers
generated from a zero-mean normal distribution with standard deviation
of unity.

The next step is to calculate $S$ for each component of ${\bf X}(0)$,
ordering them from the lowest to the highest value of $S$. Then the
first $N_{elite}$ set of parameters is selected, i.e. the
$N_{elite}$-samples with lower $S$-values, which will be labeled
hereafter as the elite sample array ${\bf X}^{elite}$.

Having determined ${\bf X}^{elite}$ at $k$th iteration, the mean
and standard deviation of the elite sample are calculated, $\bar{{\bf
    x}}^{elite}_i(k)$ and ${\bf \sigma}^{elite}_i(k)$
respectively, using:

\begin{equation}
  \bar{{\bf x}}^{elite}_i(k)=\frac{1}{N_{elite}}\sum\limits_{j=1}^{N_{elite}}X^{elite}_{ji},
\end{equation}

\begin{equation}
  {\bf \sigma}^{elite}_i(k)=\sqrt{\frac{1}{\left(N_{elite}-1\right)}\sum\limits_{j=1}^{N_{elite}}\left[X^{elite}_{ji}-\bar{{\bf x}}^{elite}_j(k)\right]^2}.
\end{equation}

In order to prevent convergence to a sub-optimal solution due to the
intrinsic rapid convergence of the CE method, \cite{kro06} suggested
the implementation of a fixed smoothing scheme for $\bar{{\bf
    x}}^{elite,s}_i(k)$ and ${\bf \sigma}^{elite,s}_i(k)$:

\begin{equation}
  \bar{{\bf x}}^{elite,s}_i(k)=\alpha^\prime\bar{{\bf 
x}}^{elite}_i(k)+\left(1-\alpha^\prime\right)\bar{{\bf x}}^{elite}_i(k-1),
\end{equation}

\begin{equation}
  {\bf \sigma}^{elite,s}_i(k)=\alpha_{d}(k){\bf 
\sigma}^{elite}_i(k)+\left[1-\alpha_{d}(k)\right]{\bf \sigma}^{elite}_i(k-1),
\end{equation}
where $\alpha^\prime$ is a smoothing constant parameter
($0<\alpha^\prime< 1$) and $\alpha_{d}(k)$ is a dynamic smoothing
parameter at $k$th iteration:

\begin{equation}
  \alpha_{d}(k)=\alpha-\alpha\left(1-k^{-1}\right)^q,
\end{equation}
with $0<\alpha< 1$ and $q$ being an integer typically between 5 and 10
(\cite{kro06}).

As mentioned before, such parametrization prevents the algorithm from
finding a non-global minimum solution since it guarantees polynomial
speed of convergence instead of exponential.

The array ${\bf X}$ at $k$th iteration is determined analogously to
equation (1):

\begin{equation}
  X_{ij}(k)=\bar{{\bf x}}^{elite,s}_i(k)+{\bf \sigma}^{elite,s}_i(k) G_{i,j},
\end{equation}

The optimization stops when either the mean value of ${\bf
  \sigma}^{elite,s}_i$ is smaller than a pre-defined value or the
maximum number of iterations $k_{max}$ is reached.

\section[]{Defining the likelihood}

To implement an objective function we use a weighted likelihood
function similar to the one proposed by \cite{FJ82} and Hernandez \&
David Valls-Gabaud (2008) to fit theoretical isochrones to star
cluster data.  In the first work the authors derive what they called
the Near Point Estimator based on a Maximum Likelihood analysis of
cluster data where the fitting statistic measures the overall
coincidence of model isochrones and observed data, assuming uniform
star distribution along the isochrone. The Near Point Estimator is
derived in \cite{FJ82} and we refer the reader to this work for the
details. The authors show that the probability of a given measurement
to be a cluster star related to a given theoretical isochrone is given
by $ln(Prob) \propto d_{min}^2$, where $d_{min}^2$ is the minimum
distance from the observed point to the model isochrone, being valid
for any number of distinct measurements. In the latter work the
authors use a similar method but take the Bayesian approach to solving
the problem by setting a likelihood function very similar to the Near
Point Estimator of the previous work. Both works relate the
probability of a given star of belonging to a given model isochrone to
distances in the CMD.

In this work we adopt a similar path to define our likelihood function
which will then be maximized by the CE method described
previously. The major difference is that we include a weighting
factor, discussed in detail below, in a semi-Bayesian approach, which
defines our prior knowledge based on the observed data-set. Unlike the
work of \cite{FJ82} we do not assume uniform distribution of stars
along the isochrone. To accomplish this we define the isochrone points
by sampling from an initial mass function (IMF), randomly generating a
number of stars in the mass range of the original isochrone. Because
we sample from a given IMF, we are also able to directly account for
binaries. We have done so assuming a binary fraction of 100\% with
companions drawn randomly from the same IMF as the cluster
stars. With these randomly generated stars we obtain a synthetic
cluster which can be compared to the observed data set through a given
metric. Because this procedure populates the CMD in the correct manner
through the IMF there is no need to introduce approximations to
account for weights along the isochrone, therefore a statistic which
sums over the distances in magnitude space of a given observed star to
the generated points will give the observed point probability of
belonging to that specific model isochrone, which is calculated by:

\begin{eqnarray}
P(V,BV,UB|I_N)_l= \sum_{m} \frac{1}{\sigma_{V_l}\sigma_{BV_l}\sigma_{UB_l}} \times \\
EXP\left [-\frac{1}{2} \left ( \frac{V_l-{I_N}_{,V_m}} {\sigma_{V_l}} \right )^2 \right ] \times \nonumber\\
EXP\left [-\frac{1}{2}\left ( \frac{BV_l-{I_N}_{,BV_m}}{\sigma_{BV_l}} \right )^2 \right ] \times \nonumber\\
EXP\left [-\frac{1}{2}\left ( \frac{UB_{l}-{I_N}_{,UB_m}}{\sigma_{UB_l}} \right )^2 \right ] \nonumber
\end{eqnarray}

where $I_N$ is a tabulated Isochrone function defined by $m$ points,
$V_l$ is the observed V magnitude, $BV_l$ and $UB_l$ the color
indexes, $\sigma_{V_l}$ is the error on the $V$ magnitude of star $l$,
$\sigma_{BV_l}$ and $\sigma_{UB_l}$ are the errors on the $BV$ and
$UB$ colors of star $l$. 

Because the $I_N$ isochrone is a discretely tabulated function, the
continuous optimization algorithm described previously was adapted to
find the nearest parameters thus introducing a grid resolution error
in the values obtained. The grid used is comprised of isochrones of
ages from $log(age)=6.6$ to $log(age)=10.15$ with a step of
$log(age)=0.05$. The uncertainty resulting from the grid resolution
has been incorporated in the final quoted errors.

The weighted likelihood function is then given in the usual manner by:
\begin{equation}
\mathcal L = \prod_{l}^{N_d}P(V,BV,UB|I_N)_l \times W_l
\end{equation}
where $W_l$ is the weight for a given star as determined from the data
using the non-parametric technique described in the following section.

 The likelihood above is used to define the objective function
  $S({\bf X})$ of the optimization algorithm as follows:

\begin{equation}
S({\bf X}) = -log(\mathcal L({\bf X})) 
\end{equation}
where ${\bf X}$ is the vector of parameters that define a given
isochrone ${\bf I_N}$ and the optimization is then done with respect
to N.

\section[]{Determining the weight function}
Before determining the weights of each observed star we must first
deal with the fact that the data is not free of contamination from
stars that do not belong to the cluster itself. The contamination is
usually from field stars that can be at different distances and have
different reddening values than that of the cluster. Therefore we must
introduce schemes to filter out contaminating stars as well as to
determine which stars are more likely to belong to the cluster in a
way that can be easily reproduced given simple and clear parameters.

\subsection[]{Magnitude cut-off}
The first step in the decontamination is to inspect the magnitude
cut-off of the observations. In many cases the observations are
clearly not complete down to the faintest magnitude as can be easily
seen in the histogram of the observed $V$ magnitude. With the $V$
magnitude histogram, obtained with a bin size of 0.5, we determine in
which magnitude it peaks and then reject stars that have magnitudes
higher than this threshold. This procedure takes care of a good part
of the contamination from faint stars, that are in fact intrinsically
brighter and further away than the cluster itself and thus show up as
faint stars in the CMD. However, this procedure is not sufficient to
remove contamination from stars that are within the completeness limit
of the observation.

The stars that are within the completeness limit of the observations
are usually the main source of contamination in the fainter (lower
mass) regions of the CMD where larger magnitude errors also contribute
to the confusion. Typically when a field is very crowded and this
contamination type is large, it can be identified as a triangular
region where the stars concentrate with high density and no clear
clustering around a main sequence. To treat these situations when
there is high density of field stars and no detectable clustering
around an isochrone main sequence we introduced a magnitude cut-off
that can be defined by the user. Since the turn-off point and the red
giant region of the open clusters are the two most important features
that constrain the determination of fundamental parameters, this
cut-off does not affect negatively their final value in most
situations.

\subsection[]{Cluster density profile}
Having eliminated the most obvious types of contamination we proceed
to estimating the weight of a given cluster star using non-parametric
techniques. The main assumption made here is that cluster stars are
concentrated in a limited region in the observed field and also in a
specific region of the CMD assuming they formed according to a single
isochrone. To determine the region in the observed field where the
cluster stars are located we use the position of the cluster center,
usually provided with the observational data set obtained from the
literature. In some cases where this information is not provided it
can be determined by obtaining a two dimensional histogram of star
positions and determining the location of the peak, provided sufficient
number of stars. 

With the position of each star measured relative to the cluster center
we obtain their radial distance (measured in pixels). We then use the
radial distances to calculate the number of stars at a given radial
distance using a histogram. The bin size of the histogram is
calculated with $bin=0.05\times MAX({\bf r})$, where ${\bf r}$ is the
vector of all star radial distances. For a given cluster $h$ bins will
determined and the density of stars as a function of radial distance
is then estimated by $\rho({\bf r}_h)=N_h/4 \pi {\bf r}_h^2$, where
$N_h$ is the number of stars in the $h$th bin. Typically the density
profile of a cluster falls off as the radius increases. Integrating
this density profile we can define a cluster radius where we find a
given percentage of the total number of stars in the field. The user
defined percentage value, which we call star fraction or $F_{star}$,
is then used to define two regions, Cluster and Field, using the
following integral:

\begin{equation}
\int_0^{R_{cluster}} \rho(r)/N_{star}~dr = F_{star}
\end{equation}
where $R_{cluster}$ is the cluster radius for the given $F_{star}$ fraction.

\subsection[]{Photometric uncertainty}

To determine the weight of each star we also need the photometric
error $\sigma_{phot}$ of the observed data. The error is defined as a
percentage and errors for each star in magnitude and color are
calculated using this factor. For the color errors we assume that the
photometric errors in each filter are independent of each other and
therefore calculate the error with the usual propagation formula.
Since few works in the literature include a full study of the errors
involved in obtaining magnitudes we adopted values that were
consistent with the ones obtained by \cite{M01}, where the author
presents error values for stars observed multiple times with the same
instrumentation. We also used the results of the filtering described
below to guide the final error value adopted, aiming for the most
efficient elimination of contamination.

\subsection[]{Non-parametric weight function}

With the Cluster and Field regions as well as the photometric errors
defined, the weight for each star is then estimated by comparing the
characteristics of the stars in an area around the $lth$ star in the
CMD defined by a box with dimensions $3\sigma_{V_{l}}$ by
$3\sigma_{(B-V)_{l}}$. We then calculate the average and standard
deviation of $V$ and $(B-V)$ for the stars that fall within this box
and belong to the Cluster region defined earlier. The assumption made
here is that this statistic provides the most likely position for a
cluster star in the CMD region defined by the error box of the $lth$
observed star. This is clearly not the case when there is no
detectable concentration of cluster stars relative to field stars
within the error box. This situation is more likely to happen in
clusters with heavy field contamination and large magnitude errors as
in the lower magnitude regions of the CMD. However, as discussed
earlier the magnitude cut-off can resolve this issue by eliminating
these stars from the sample to be fitted. We then calculate the weight
for the $l$th star with the expression:

\begin{eqnarray}
W_{l} = \frac{1}{\sigma_{V_{l}}\sigma_{BV_{l}}\sigma_{UB_{l}}} \times
EXP^{\frac {-(V_l-\overline{V_c})^2} {2\sigma^2_{V_c}} } \times \nonumber\\
EXP^{\frac{-(BV_l-\overline{BV_c})^2} {2\sigma^2_{BV_c}} }\times \nonumber\\
EXP^{\frac{-(UB_l-\overline{UB_c})^2} {2\sigma^2_{UB_c}} }\times \nonumber\\
EXP^{\frac{-r_l^2} {2 \left ( \frac{R_{cluster}} {3} \right ) ^2} }
\end{eqnarray}
 where $V_{l}$, $BV_{l}$ and $UB_{l}$ are the observed V
magnitude and the color indexes of the $lth$ star, $\overline{V_{c}}$,
$\overline{BV_{c}}$ and $\overline{UB_{c}}$ are the average V
magnitude and the average color indexes of the stars that fall within
the $3\sigma$ error box and belong to the Cluster region as defined
earlier. Note that according to this procedure stars that fall outside
of the Cluster region are automatically given $W_{l}=0$.

In Fig. 1 we show the results of the decontamination and weighting
process for the cluster \object{NGC~2477} and data set of \cite{Kassis1997}.
The black dots in the left graph are the selected stars after
decontamination, the open circles are stars that fall outside the
defined cluster radius (which are then eliminated from the sample to
be fitted) and light dots are stars for which no statistic was
available due to low numbers. The right graph shows open circles with
sizes scaled to their weights, with larger sizes meaning larger
weights. It is clear from these graphs that the decontamination scheme
we have adopted is not perfect and in regions where field stars have
large densities in the CMD non-cluster stars are likely to survive the
process. However, as we can see in the right graph, the weighting
procedure does a good job of assigning low values to these fields
stars. In any case, as discussed before, we have introduced a magnitude
cut-off to eliminate these regions altogether from the fitting
process if necessary.

\subsection[]{Implementing the Cross entropy algorithm}
In our problem $I_N$ is a tabulated function taken from a grid of models
calculated by Padova database of stellar evolutionary tracks and
isochrones \cite{Girardi2000, Marigo2008}.

\begin{figure*} \begin{center}
    \includegraphics[scale=0.5]{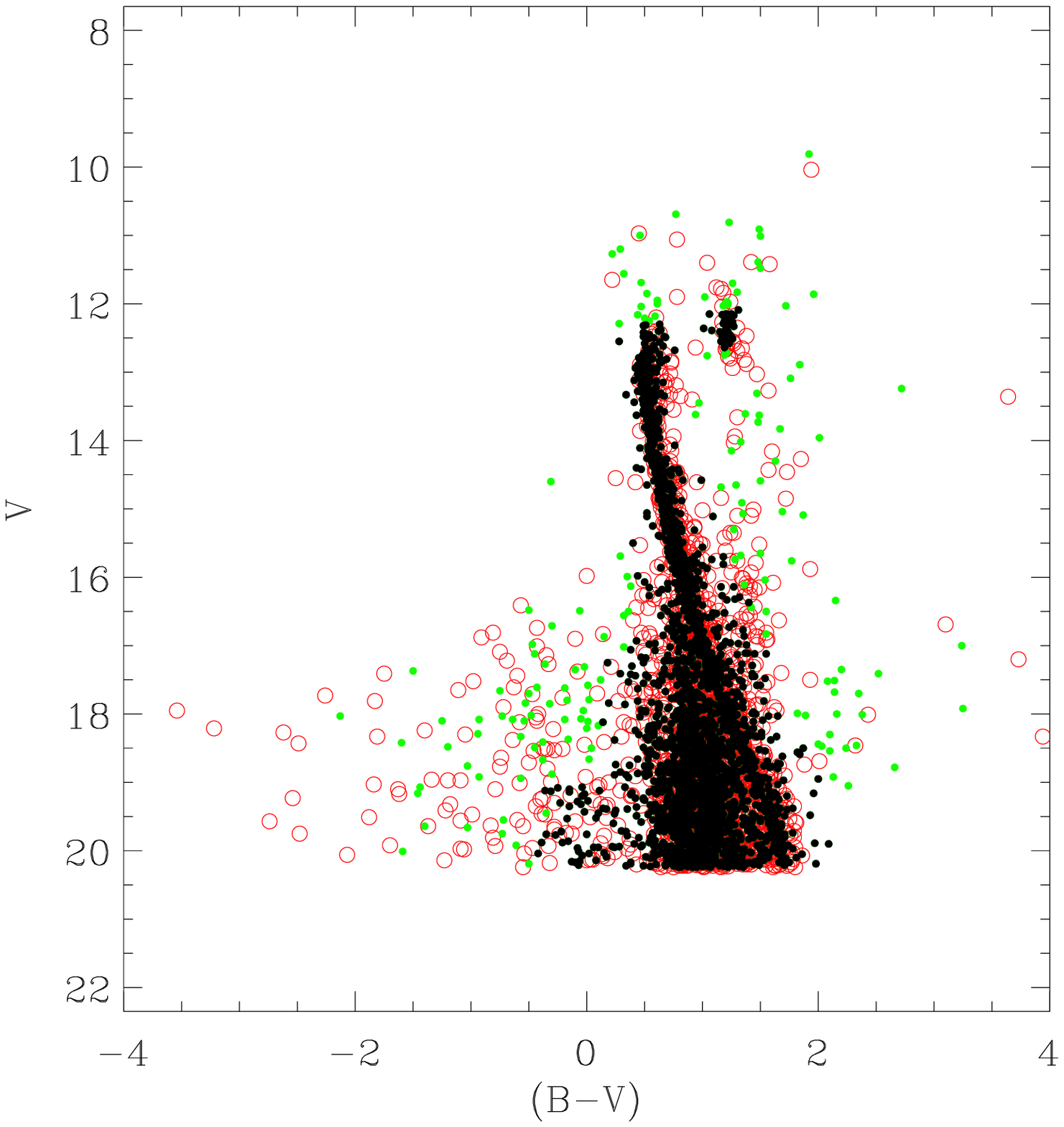}
    \includegraphics[scale=0.5]{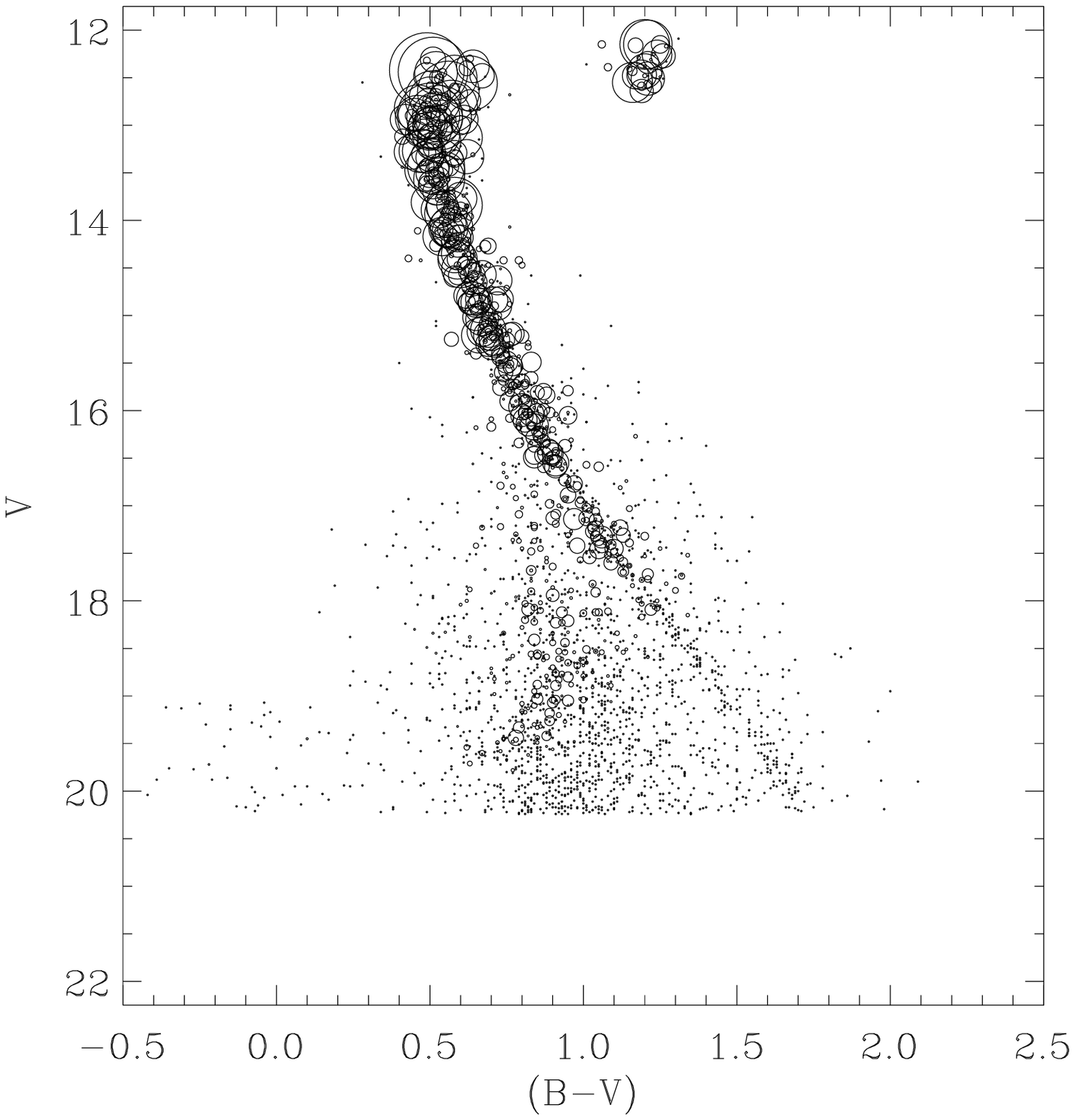} 
\end{center}

\caption{Result of the decontamination (left) and weighting process
  (right) for the cluster \object{NGC~2477}. The black dots in the left graph
  are the selected stars after decontamination, the open circles are
  stars that fall outside the defined cluster radius and light dots
  are stars for which no statistic was available due to low
  numbers. The right graph shows open circles with sizes scaled to
  their weights, with larger sizes meaning larger weights.}

\end{figure*}

The tabulated isochrones are defined by 2 parameters, namely, age and
metallicity and to compare to observed data we also need distance and
extinction, which we consider constant. These are the parameters we
wish to optimize thus fitting the tabulated isochrone to the observed
data. The parameter ranges for generating ${\bf X}$ are pre-defined by
the user and should be representative of the problem being
optimized. In general, the CE algorithm is very forgiving of large
parameter spaces, being very efficient in quickly zoning in on optimal
regions. In the isochrone fitting done in this work we defined the
parameter space as follows:

\begin{enumerate}
\item {\bf Age}: from $log(age)=6.60$ to $log(age)=10.15$ encompassing the full
  range of theoretical isochrones;
  \item {\bf distance}: from 1 to 10000 parsecs
  \item ${\bf E(B-V)}$: from 0.0 to 3.0
  \item {\bf Z}: fixed (used literature values)
\end{enumerate}

The algorithm has been written to allow the optimization of the
metallicity as well and work is under way to fully implement this
feature. However, since we are mainly concerned with benchmarking the
method and comparing results to the literature values provided by
\cite{PN06} we have kept it fixed to values obtained from the
literature.

\begin{table*}
  \caption{Results for synthetic clusters studied by the fitting method.}
\centering
\begin{tabular}{lcccccc}
  \hline\hline
  {Cluster} & {$N_{stars}$} & {Contamination (\%)} & {$3\sigma_{phot}$(\%)} & {$E(B-V)$(mag)} & {Distance
    (pc)} & {$log(Age)$ (yr)} \\
  \hline
  SC~01     & 432 & 0\%  & 1.0  & $0.40 \pm 0.01$ & $2112 \pm 51$  & $8.65 \pm 0.05$  \\  
  SC~02     & 480 & 20\% & 1.0  & $0.40 \pm 0.01$ & $2062 \pm 43$  & $8.71 \pm 0.05$  \\ 
  SC~03     & 444 & 50\% & 1.0  & $0.38 \pm 0.01$ & $2073 \pm 30$  & $8.70 \pm 0.07$  \\ 
  SC~04     & 65  & 0\%  & 1.0  & $0.38 \pm 0.02$ & $2008 \pm 94$  & $8.70 \pm 0.07$  \\
  SC~05     & 113 & 20\% & 1.0  & $0.40 \pm 0.03$ & $2060 \pm 60$  & $8.75 \pm 0.06$  \\
  SC~06     & 61  & 50\% & 1.0  & $0.40 \pm 0.02$ & $2102 \pm 58$  & $8.70 \pm 0.07$  \\
  \hline
\end{tabular}

\tablefoot{Synthetic clusters were generated with parameters log(age)=8.70 yr, distance=2100 pc, 
  E(B-V)=0.40, Z=0.019) and $N_{stars}$, including the given contamination fraction 
  and photometric accuracy of $3\sigma_{phot}$.}

\label{tab1}
\end{table*}


The filtered sample is then fed through the optimization algorithm
described previously which then minimizes the objective function ${\bf
  S}$ thus maximizing the likelihood function to find the best fitting
values for age, distance and reddening for a given metallicity.

An important point to be made is that although in the algorithm all
parameters can be fit simultaneously, we opted to take the usual
procedure of determining the $E(B-V)$ parameter first, using only the
colour-colour diagram and the ZAMS, and then performing the fit for
the other parameters in the CMD. However, to ensure that the fit has
some liberty in accommodating other possibilities in the second stage
of the fitting we allow $E(B-V)$ to vary in a range of $10\%$ of the
value determined in the first step.

An advantage of this fitting procedure is that it allows for
determination of parameter errors through Monte-Carlo techniques. To
accomplish this we perform the fit for each data set $N_{Run}$ times,
each time re-sampling from the original data set with replacement to
perform a bootstrap procedure as well as generating new isochrone
points from the adopted IMF as described previously. For each run we
also replace the stars chosen in the new bootstrap sample with ones
obtained by randomly generating values of $(U,B,V)_l$ drawn from a
normal distribution centered at the original data value and with
$\sigma=\sigma_{phot}$. The final uncertainties in each parameter are
obtained by calculating the standard deviation of the $N_{Run}$ fit
values.

\section{Validating the method}

\begin{figure*}
\begin{center}
\includegraphics[]{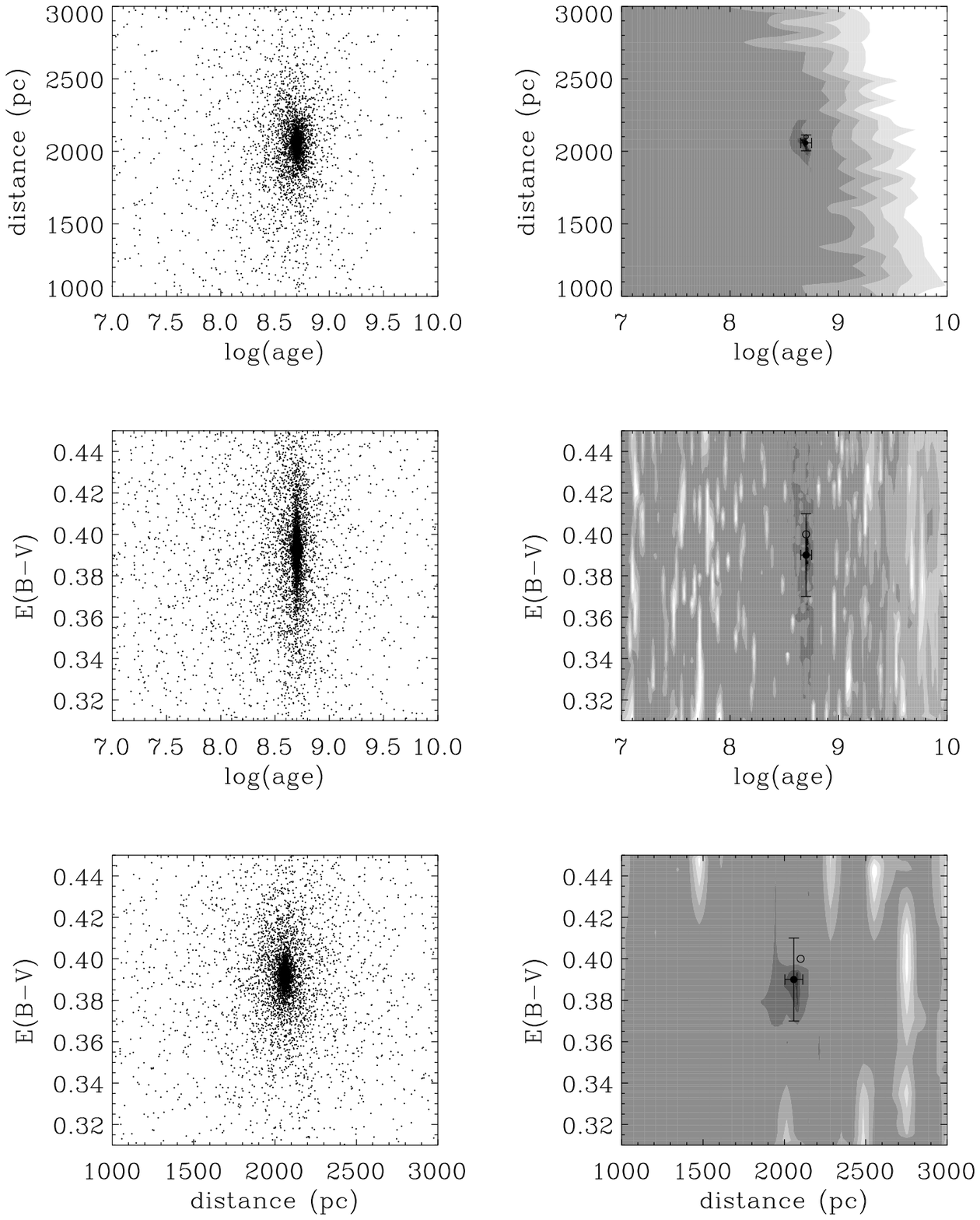}
\end{center}
\caption{Parameter space coverage of a full run of the optimization
  algorithm (left column) as well as the 2D likelihoods (right column)
  for the fitted parameters for synthetic cluster SC 06. In the right
  column open circles represent the correct solution and filled
  circles the final parameter value obtained from the algorithm.}
\end{figure*}

To validate the method we applied the fitting algorithm to a set of
synthetic clusters generated from a predefined isochrone of a given
age and metallicity. We set the distance and reddening and generate a
set of stars drawn from a probability distribution obtained from a
Salpeter IMF. To make the test more realistic we also introduce some
contaminating field stars, generated in a bigger volume of space and
with varying reddening constants determined randomly. It is important
to point out that the synthetic cluster is not supposed to be a
realistic rendition of a real open cluster but just a tool to gage the
capability of the method to recover well defined parameters. In this
case we know precisely the age and metallicity of the isochrone as
well as the IMF that generated the stars. In real situations the
contaminations can come from many sources and have very different
characteristics. We did not run exhaustive tests with synthetic
clusters and different contamination schemes to determine their effect
in the results. However as the results will show in the following
sections the filtering scheme explained previously does a good job in
rejecting stars with low probability of belonging to the cluster.

\begin{figure*} \begin{center}
    \includegraphics[scale=0.6]{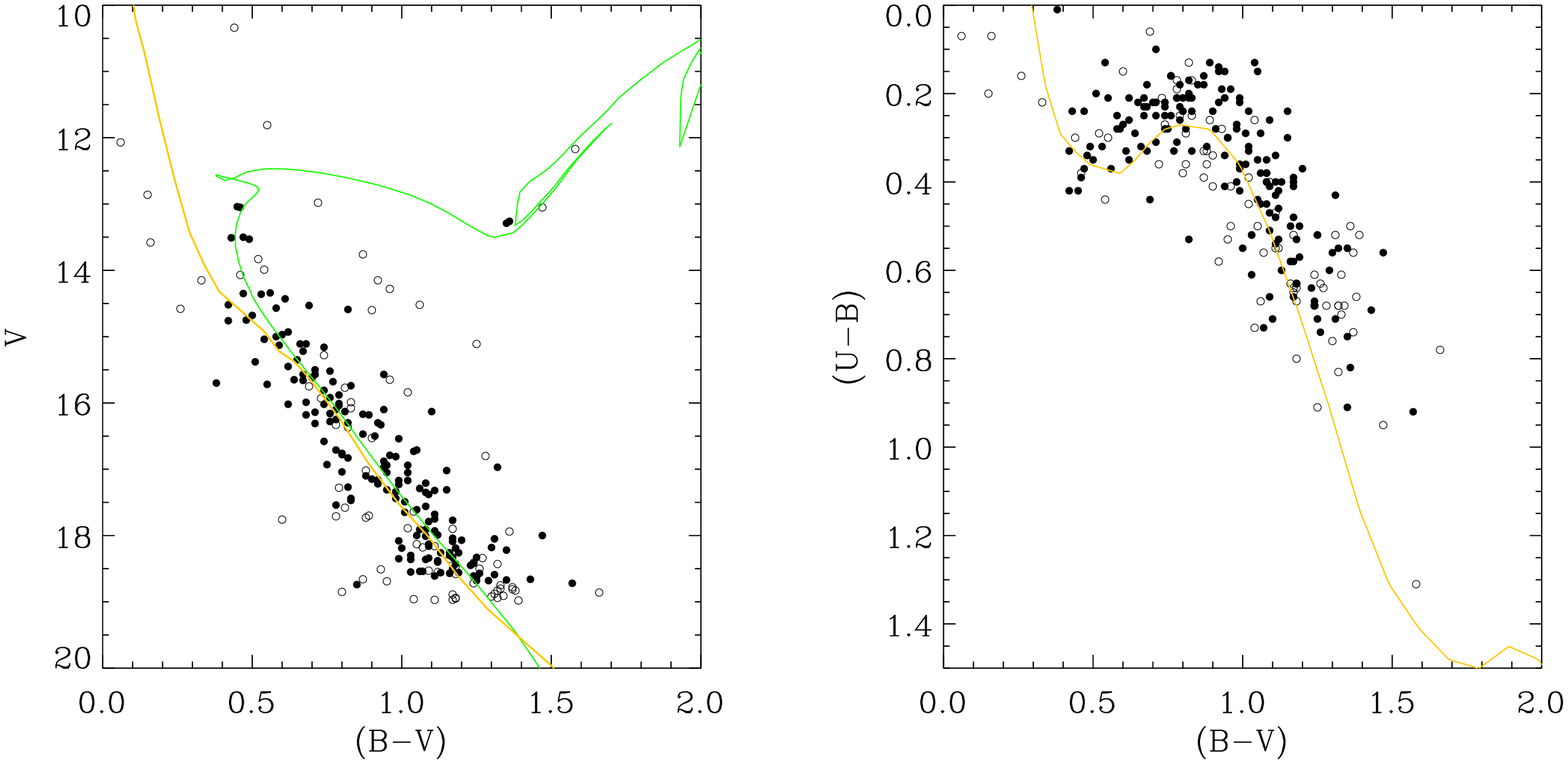} 
\end{center}

  \caption{Result of the fitting method for the synthetic cluster
    SC~06 with the ZAMS (thin line) and the fitted isochrone (thick
    line) where we show the rejected stars by our filtering method (open circles) 
    as well as stars used
    in the fit (filled circles)}

\end{figure*}

\begin{figure*} \begin{center}
    \includegraphics[scale=0.6]{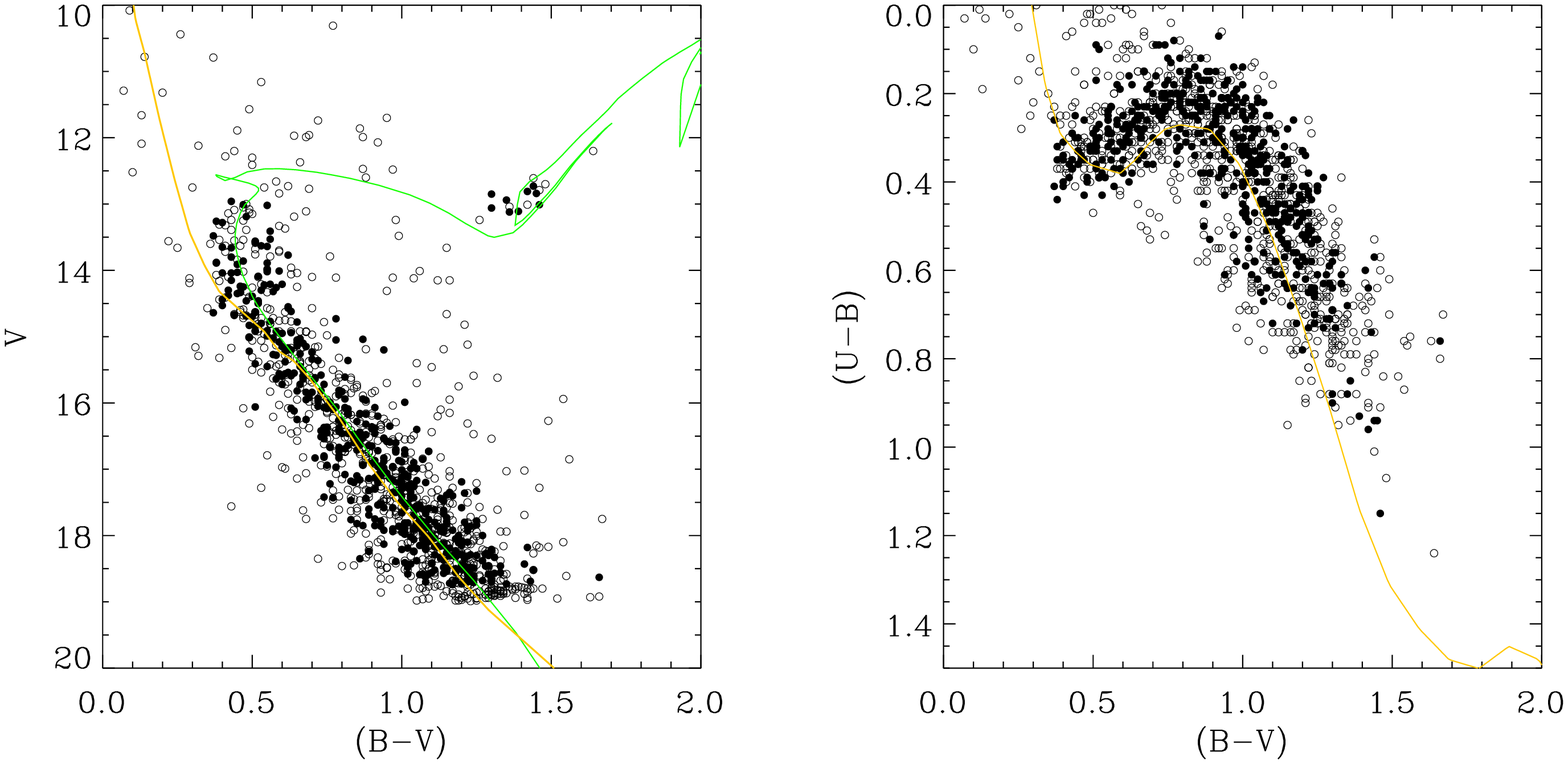} 
\end{center}

  \caption{Result of the fitting method for the synthetic cluster
    SC~03 with the ZAMS (thin line) and the fitted isochrone (thick
    line) where we show the rejected stars by our filtering method (open circles) 
    as well as stars used
    in the fit (filled circles)}

\end{figure*}

\subsection{Synthetic cluster}

Based on other works published in the literature, we generated
synthetic clusters with typical numbers of stars, photometric errors
and contamination. In all tests we were able to recover the parameters
used to create the cluster within reasonable errors considering the
level of contamination and photometric errors as well as the number of
generated stars. As would be expected, the method is sensitive to the
number of observed stars and the photometric errors, but in all cases
gave good values within the errors established. In general, as long as
the cluster was well sampled, i.e. had enough member stars, the method
converged well. Poorly populated clusters should always be treated
with care or be avoided by this type of technique. Even so, we managed
to fit synthetic clusters with as little as 20 stars with errors in
the order of 25\%. The actual accuracy will depend on what section of
the isochrone is sampled, especially for the age, since the presence
of red giants usually provide a strong constraint.

The synthetic clusters were generated with the following parameters:

\begin{itemize}
\item ${\bf log(age)}=8.70~yr$
\item {\bf distance}=2100 pc
\item ${\bf E(B-V)}$=0.40
\item {\bf Z}=0.019
\item Number of stars $\approx$ 100 to 1000
\item Contamination 20\% to 50\%
\item Photometric error $3\sigma_{phot}=1\%$
\end{itemize}

The final results obtained for all synthetic clusters using the CE
fitting technique are summarized in Table 1. The tunning parameters
that gave consistent convergence to the correct answer in all tested
cases were $\alpha=0.6$, $q=0$, $N_{elite}=50$, a sample of $N=500$
trial solutions per iteration, with a maximum number of 20
iterations. We used the tuning parameter values listed previously in
all fits in this work.

In Fig. 2 we show plots of the parameter space coverage of a full run
of the optimization algorithm as well as the 2D likelihood for the
fitted parameters for synthetic cluster SC01. It is clear in these
plots that the method finds the optimal solution even in considerably
irregular likelihood space. The error bars shown were obtained from
the bootstrap procedure. Note that the solution obtained from the
fitting algorithm (filled circles) are consistent with the values used
to generate the synthetic cluster (open circles). The non-coincidence
of the open circle positions and the likelihood maxima is due to the
generation of synthetic clusters with limited number os stars as well
as the errors used. The smaller the error and the greater the number
of stars the closer the positions would be.

The generated cluster data as well as the final fitting isochrone is
shown for two of the synthetic clusters (SC03 and SC06) in Fig. 3 and
Fig. 4.

To quantify the effect of adopting 100\% binaries in the fitting
procedure we generated a synthetic cluster using the same parameters
adopted for the previous tests but now with a binary fraction of
50\%. We then performed fits using different binary fractions of 0\%,
25\%, 50\%, 75\% and 100\%. The results are shown in Fig. 5. The main
effect observed is the overestimation of the distance and reddening
for adopted binary fractions that are higher than the true
value. However, as Fig. 5 shows, in the most extreme case this
introduces a systematic error of about 10\% in the determined distance
and reddening and less than 1\% for the age. Ideally the binary
fraction should be determined through other means and then
incorporated in the fit performed. Adopting a binary fraction of 100\%
essentialy means that the results should be viewed as upper limits for
distance and reddening.

We also probed the effect of using different IMF
parametrization. Using a synthetic cluster with no binaries we
obtained fits using IMF exponents of 1.0, 1.5, 2.0, 2.1, 2.2 and
2.35. The results are shown in Fig. 6 where it is clear that the IMF
variations do not affect the final results significantly.

\begin{figure}
    \includegraphics[width=\columnwidth]{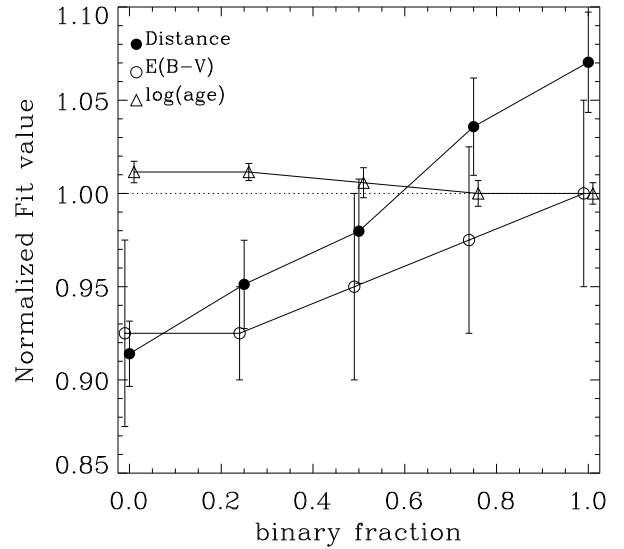} 
    \caption{Results of the fitting method for one realization of a
      synthetic cluster with a binary fraction of 0.5. The final fit
      values obtained were normalized to the expected (true) values
      (as defined in sec. 5.1). The points have been slightly shifted
      in the x axis for clarity.}

\end{figure}

\begin{figure}
    \includegraphics[width=\columnwidth]{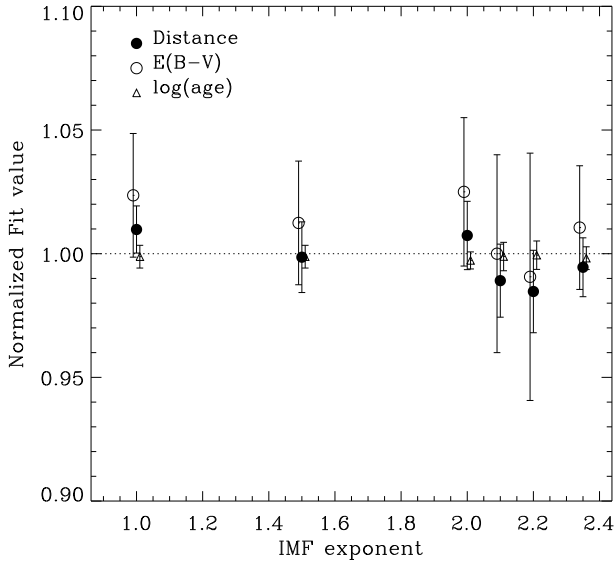} 
    \caption{Results of the fitting method for one realization of a
      synthetic cluster using different IMF exponents. The final fit
      values obtained were normalized to the expected (true) values
      (as defined in sec. 5.1). The points have been slightly shifted
      in the x axis for clarity.}

\end{figure}

\section{Fitting published data}

\begin{table}
  \caption{Cross-Entropy fit parameters} 
\begin{tabular}{lcccc}
\hline\hline
{Cluster} & {$V_{cut}$} & {${\bf F}_{star}$} & {$3\sigma_{phot}$} & {Ref.} \\
        &  {$mag$}     &              {$(\%)$} & {$(\%)$}  & \\
\hline

\object{NGC~2477}     &$17.00 $& $95$ &$ 1.0  $ & 152   \\ 

\object{NGC~2477}     &$17.00 $& $95$ &$ 1.0  $ & 152   \\ 
\object{NGC~7044}     &$21.75 $& $95$ &$ 1.0  $ &  62   \\
\object{NGC~2266}     &$18.75 $& $95$ &$ 1.5 $ & 41    \\
\object{Berkeley 32}  &$20.25 $& $95$ &$ 1.5  $ & 40    \\
\object{NGC~2682}     &$18.00 $& $98$ &$ 1.5 $ & 335   \\
             &$18.00 $& $99$ &$ 1.5 $ & 31    \\
             &$18.00 $& $98$ &$ 1.5 $ & 54    \\
\object{NGC~2506}     &$18.00 $& $98$ &$ 1.5  $ & 284   \\
             &$17.75 $& $98$ &$ 1.5 $ & 163   \\
\object{NGC~2355}     &$19.75 $& $98$ &$ 1.0  $ & 217   \\
             &$18.25 $& $98$ &$ 1.0  $ & 44    \\
\object{Melotte 105}  &$16.75 $& $98$ &$ 1.5  $ & 289   \\
             &$16.75 $& $98$ &$ 1.5  $ & 32    \\
\object{Trumpler 1}         &$19.00 $& $85$ &$ 1.5  $ & 320   \\ 
             &$19.00 $& $95$ &$ 1.5 $ & 86    \\
\hline
\end{tabular}
\tablefoot{here $V_{cut}$ is the adopted cut-off in magnitude in V, $\rho_{star}$ is the density of stars in the observed field as a function of radius determined from the cluster
  center and $3\sigma_{phot}$ is the photometric error as described in  the text.}
\begin{flushleft} 
\tiny
References: \\
152 =  \cite{Kassis1997} \\
62  =  \cite{Aparicio1993}\\ 
41  =   \cite{Kaluzny1991}  \\
40  =  \cite{Kaluzny1991} \\ 
335 =  \cite{Henden2003}  \\
31  =  \cite{Gilliland1991} \\
54  =  \cite{Montgomery1993}(adopted logt = 9.6)\\
284 =  \cite{Kim2001} (adopted mean values: see table 5 of the paper) \\    
163 =  \cite{Marconi1997}\\  
217 =  \cite{Ann1999}  \\
44  =  \cite{Kaluzny1991}\\   
289 =  \cite{Sagar2001}\\
32  =  \cite{Kjeldsen1991}  \\
320 =  \cite{Yadav2002}\\ 
86  =  \cite{Phelps1994}\\
\end{flushleft}
\end{table}

Having validated the fitting technique with the synthetic clusters we
proceeded to the fitting of real cluster data. We selected a set of
well studied clusters based on the work of \cite{PN06} where the
authors performed a statistical analysis of the determined physical
parameters of open clusters published in the literature to
characterize the current status of knowledge and the precision of open
cluster parameters such as age, reddening and distance. The authors
defined a sample of 72 open clusters with the most precise known
parameters, which they suggest, should be used as standards for future
theoretical work.

In our work we pre-selected 29 clusters from the standard sample
proposed by \cite{PN06} that had at least 5 independent fundamental
parameter estimates to be representative of the heterogeneity of the
published results. This allowed us to compare our results to others
obtained in the literature considering a more representative estimate
of the errors in each parameter.

The observational data was obtained through the WEBDA
catalog\footnote{available at http://obswww.unige.ch/webda}
(\cite{Mermilliod1995}) for each cluster. We only selected clusters with
U band photometry as this allowed us to determine the reddening
through the color-color diagram of open clusters as traditionally done
(see \cite{Phelps1994}). We did not use data sets that had mixed
observations from different authors, using only those sets that
originated from a single source. For the 29 previously selected
clusters only 13 satisfied these conditions. For some of the clusters
we also obtained multiple data sets that satisfied our criteria and
those were also used. From the 13 final clusters we selected a sample
of 9 to which we applied our fitting technique.

In Table 2 we present the fit parameters used and in Table 3 we
present the final fitting results as well as the reference numbers (as
defined in the WEBDA catalog) of each cluster studied using our
method.

As mentioned before, since the selected data samples did not contain
photometric error estimates in WEBDA, we have adopted values based on
the results of the filtering algorithm applied. The photometric errors
presented in Table 2 are those that gave the best result for the
filtering method based on elimination of obvious contaminating stars
and good definition of the turn-off and main sequence of the
cluster. Because of this procedure these estimated errors should not
be taken as formal photometric errors for the data.

\begin{figure}
\begin{center}
\includegraphics[width=\columnwidth]{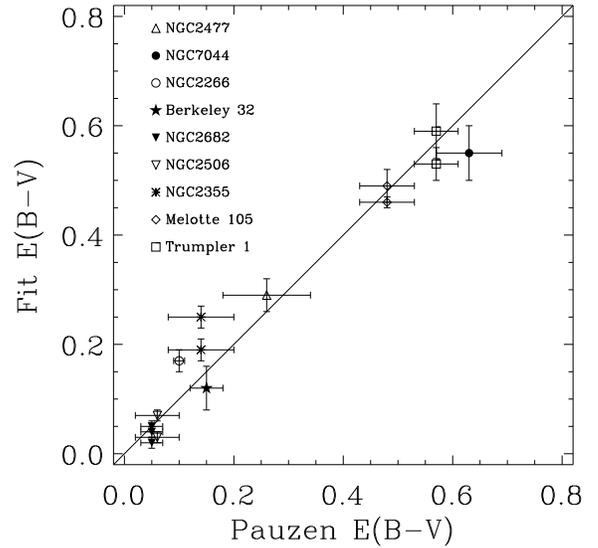}
\end{center}
\caption{Comparison of our fit results to those of \cite{PN06} for
  the parameter $E(B-V)$. The error bars are obtained from the errors
  presented in Table 3 for our fit and literature respectively.}
\end{figure}

\begin{figure}
\begin{center}
\includegraphics[width=\columnwidth]{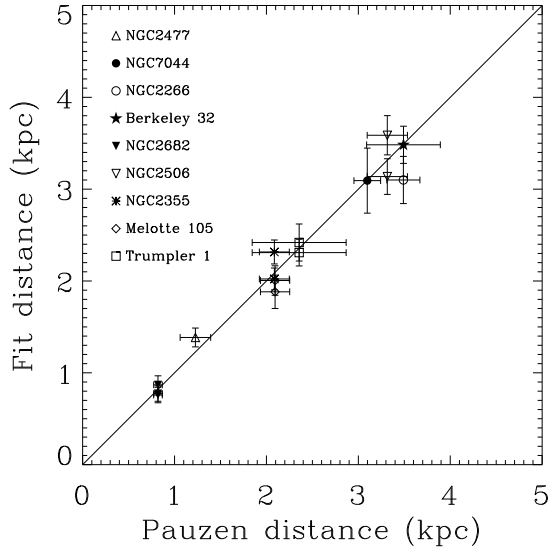}
\end{center}
\caption{Same as in Fig. 4 for the parameter distance.}
\end{figure}

\begin{figure}
\begin{center}
\includegraphics[width=\columnwidth]{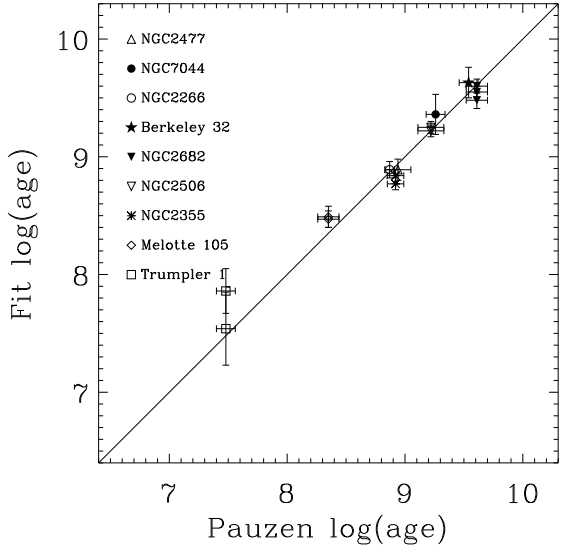}
\end{center}
\caption{Same as in Fig. 4 for the parameter $log(age)$.}
\end{figure}

\section{Results and discussion}

It is difficult to compare our results directly to other individual
results in the literature given that they use different methods to
analise the CMD as well as different isochrone models in some
cases. For this reason we opted to use the \cite{PN06} values instead
of individual parameter determinations as we feel these are more
representative of real values and the errors reflect the precision of
all techniques used so far. In this way we avoid any potential bias to
any one given technique or observational data set and heterogeneity
can be accommodated by the quoted errors, also allowing us a more
quantifiable comparison to other parameter determinations.

We present the comparison of our results to those given by
\cite{PN06} for the 9 open clusters studied in Figures 7 to 9. The
average and standard deviation of the differences of our
results to those of \cite{PN06} are:

\begin{itemize}
\item ${\bf E(B-V)}=-0.005\pm 0.047 ~mag$
\item ${\bf distance}=31\pm 182 ~pc$
\item ${\bf log(age)}= -0.03\pm 0.13 ~ yr$
\end{itemize}

The differences between \cite{PN06} and our work are small except in 4
cases where significant discrepancies where found and are thus
discussed in more detail below. The uncertainties of our method are
also in general smaller than the ones quoted by \cite{PN06}. The
difference is to be expected since our uncertainties are mainly
related to the fitting procedure and the data set used while the
values of \cite{PN06} are the standard deviation over multiple
independent estimations using possibly very different methods and
models. It is important to point out that although the clusters
selected for this work are taken from a list considered to be of
standards, there is no absolute guarantee that their results are the
correct ones and this should be kept in mind when comparing parameter
values.

\begin{table*}
  \caption{Basic parameters obtained for the investigated  clusters.} 
  \begin{tabular}{lccccccccc}
    \hline\hline
              & \multicolumn{4}{c}{Fit}                      & &\multicolumn{3}{c}{Literature} \\
    {Cluster} &   {$E(B-V)$} & {Distance} & {Log(Age)} & {Z} & & {$E(B-V)$} &  {Distance} & {Log(Age)}   &  {Ref.} \\
              &      (mag)        &     (pc)   &     (yr)   &     & &    (mag)        &      (pc)   &       (yr)   &          \\  
    \hline 
    \object{NGC~2477}       & $0.29\pm0.03$ & $1385\pm64  $  & $8.90\pm0.09$  & $0.019$ &  & $0.26\pm0.08$    &  $ 1227\pm166$    & $8.94\pm0.11 $ & 152   \\
    \object{NGC~7044}       & $0.55\pm0.05$ & $3093\pm345 $  & $9.35\pm0.17$  & $0.019$ &  & $0.63\pm0.06$    &  $ 3097\pm145$    & $9.26\pm0.08 $ &  62   \\
    \object{NGC~2266}       & $0.17\pm0.02$ & $3100\pm244  $ & $8.90\pm0.07$  & $0.008$ &  & $0.10\pm0.01$    &  $ 3490\pm180$    & $8.87\pm0.04 $ & 41    \\
    \object{Berkeley 32}    & $0.12\pm0.04$ & $3483\pm186  $ & $9.65\pm0.13$  & $0.008$ &  & $0.15\pm0.03$    &  $ 3491\pm401$    & $9.54\pm0.08 $ & 40    \\
    \object{NGC~2682}       & $0.02\pm0.01$ & $774\pm25  $   & $9.55\pm0.05$  & $0.019$ &  & $0.05\pm0.02$    &  $  820\pm47$     & $9.61\pm0.09 $ & 335   \\
                   & $0.05\pm0.01$ & $758\pm26   $  & $9.60\pm0.06$  & $0.019$ &  & $0.05\pm0.02$    &  $  820\pm47$     & $9.61\pm0.09 $ & 31    \\
                   & $0.04\pm0.01$ & $869\pm57   $  & $9.50\pm0.07$  & $0.019$ &  & $0.05\pm0.02$    &  $  820\pm47$     & $9.61\pm0.09 $ & 54    \\
    \object{NGC~2506}       & $0.03\pm0.01$ & $3587\pm198 $  & $9.20\pm0.05$  & $0.008$ &  & $0.06\pm0.04$    &  $ 3315\pm219$    & $9.22\pm0.11 $ & 284   \\
                   & $0.07\pm0.01$ & $3137\pm177  $ & $9.25\pm0.05$  & $0.008$ &  & $0.06\pm0.04$    &  $ 3315\pm219$    & $9.22\pm0.11 $ & 163   \\
    \object{NGC~2355}       & $0.25\pm0.02$ & $2316\pm103  $ & $8.80\pm0.05$  & $0.008$ &  & $0.14\pm0.06$    &  $ 2086\pm163$    & $8.92\pm0.07 $ & 217   \\
                   & $0.19\pm0.02$ & $2022\pm88  $  & $8.85\pm0.05$  & $0.008$ &  & $0.14\pm0.06$    &  $ 2086\pm163$    & $8.92\pm0.07 $ & 44    \\
    \object{Melotte 105}    & $0.47\pm0.02$ & $1750\pm111 $  & $8.40\pm0.06$  & $0.019$ &  & $0.48\pm0.05$    &  $ 2094\pm159$    & $8.35\pm0.09 $ & 289   \\
                   & $0.49\pm0.03$ & $2005\pm139 $  & $8.45\pm0.07$  & $0.019$ &  & $0.48\pm0.05$    &  $ 2094\pm159$    & $8.35\pm0.09 $ & 32    \\
    \object{Trumpler 1}           & $0.59\pm0.05$ & $2419\pm185 $  & $7.85\pm0.19$  & $0.019$ &  & $0.57\pm0.04$    &  $ 2356\pm511$    & $7.48\pm0.08 $ & 320   \\
                   & $0.53\pm0.03$ & $2309\pm121 $  & $7.55\pm0.31$  & $0.019$ &  & $0.57\pm0.04$    &  $ 2356\pm511$    & $7.48\pm0.08 $ & 86    \\
    \hline      
  \end{tabular} 
  \tablefoot{The numbers, 
    in the last column are the WEBDA reference codes, $E(B-V)$ is the extinction, {\em d} the distance to the cluster, 
    $\log(Age)$ the logarithm of the age (in years), Z the adopted metallicity.
    The literature values are those of \cite{PN06}.}
\end{table*}

 \subsection{\object{Melotte 105}}
 For the open cluster Melotte 105 we find a discrepancy in the
 distance determined using the data from \cite{Sagar2001} when
 compared to the result obtained by \cite{PN06}. This can be explained
 by the stars used in our fit as result of the filtering technique as
 well as the weights assigned to them (see Figs 17 and 18). It is
 clear from the figures that the data from \cite{Sagar2001} is
 somewhat different from the one obtained by \cite{Kjeldsen1991}. In
 this case it is difficult to determine which is the best result
 without more observational information. Given that other results were
 obtained through visual fits we cannot compare each weight
 attribution or determine which stars were deemed more important by
 the authors, however it seems likely that \cite{Sagar2001} gave more
 importance to stars around V=11.5 that in our case were eliminated by
 the decontamination.

 \subsection{\object{NGC 2355}} 
 
 For the open cluster \object{NGC 2355} we find significant differences in
 $E(B-V)$, distance and $log(Age)$ for the fit results using the
 \cite{Ann1999} (Ref. 217) data set and a small discrepancy for
 $E(B-V)$ with the data set of \cite{Kaluzny1991} when compared to the
 results obtained by \cite{PN06}. It is likely that the most important
 factor in explaining these differences is the adopted
 metallicities. For \object{NGC 2355} \cite{Kaluzny1991} adopt $Z\approx0.03$
 while we use the value of $Z\approx0.019$ $(Fe/H = -0.07 \pm 0.11)$
 obtained by \cite{Soubiran2000} through spectroscopic observations of
 24 stars, which we believe to be more reliable.

 Other factors as the obvious difference in detection efficiency of
 each observational data set as well as precision of photometry will
 affect the results. The fit parameters obtained from the data of
 \cite{Kaluzny1991} (Ref. 44) are in better agreement with the ones
 obtained by \cite{PN06}. We believe that data from \cite{Kaluzny1991}
 is more precise although not going as deep as the data from
 \cite{Ann1999} (Ref. 217), indicating that the bigger discrepancies,
 when compared to the results of \cite{PN06}, are likely due to lower
 photometric precision and higher level of contamination in the
 data. We point out that the discrepancies in the determination of
 $E(B-V)$ may be due to differences in photometry. In Fig. 16 and
 Fig. 17 one can easily see in the colour-colour diagrams for each
 data set of \object{NGC 2355} that there is a systematic difference of about
 0.1 in $(U-B)$ between the data sets.

 \subsection{\object{NGC 7044}}
 
 For the open cluster \object{NGC 7044} we find differences between our fit
 value of $E(B-V)$ to the one determined by \cite{PN06} although the
 $log(Age)$ and distance values are within the uncertainties. The
 $E(B-V)$ value of \cite{PN06} is clearly not adequate for this data
 set as can be seen in the colour-colour diagram plot of Fig. 8,
 however our value for $E(B-V)$ as well as $log(Age)$ is consistent
 with the determination of \cite{Aparicio1993} (Ref. 62), the source
 of the data.

 The large spread in the main sequence region is also an important
 factor as pointed out by \cite{Aparicio1993}, where they argue that
 the large spread may be an indication of a large number binaries and
 based on this they estimate a value of $22\%$ for the binary
 fraction. It is clear from Fig. 8 that our method of including
 binaries does a good job in accounting for their presence as can be
 seen also by the agreement in the distance determined by the fit.

 \subsection{\object{Trumpler 1}} 
  
 For the open cluster Trumpler 1 the fitted distance agrees within the
 errors to those determined by \cite{PN06} except for the
 $log(Age)$. The age difference, especially for \cite{Yadav2002}
 (Ref. 320), is likely due to the bright stars considered in the fit.

 As we can see in Fig. 21 for the data from \cite{Yadav2002}
 (Ref. 320), our filtering technique has removed some bright stars
 that may belong to the turn-off of the cluster. The removal happened
 because in all cases these were single events in that region of the
 CMD and thus no statistic could be performed. We have introduced an
 option to keep these single points if the case may present itself but
 did not use it in this cluster as it introduced considerable
 contamination along with the wanted stars in the tip of the
 turn-off. It may be the case that in these situations our errors are
 underestimated since these stars have low photometric errors when
 compared to the fainter and more numerous stars. 

 In situations like these other data may resolve the issue if
 incorporated in the calculation of the weights. One example is the
 determination of proper motions for cluster stars, as done by
 \cite{Loktin2003} for Trumpler 1, where the author determines the
 proper motion of the cluster from measurements of 12 stars. However,
 the author does not provide the individual data for the stars and
 therefore it could not be used to assign weights to the observations.
 
 The situation is similar when we examine the results presented in
 Fig. 21 for the fit of the observational data provided by
 \cite{Phelps1994} (Ref. 86). In this case our filtering technique
 also rejected single points in the CMD in the region of the
 turn-off. The same considerations made above about the errors apply
 in this case.

 As a final note we point out that significant differences exist
 between each data set. The observations of \cite{Phelps1994} have a
 limiting magnitude of $V\approx21.5$ while \cite{Yadav2002} have a
 limiting magnitude of $V\approx19.6$ yielding 1291 and 670 detected
 stars respectively both with considerable contamination from field
 stars. \cite{Yadav2002} mention in their work that they suspect the
 data from \cite{Phelps1994} suffers from calibration problems. All
 these factors are likely to be sources of the differences in the
 results we obtained.

\begin{figure*}
\begin{center}
\includegraphics[scale=0.6]{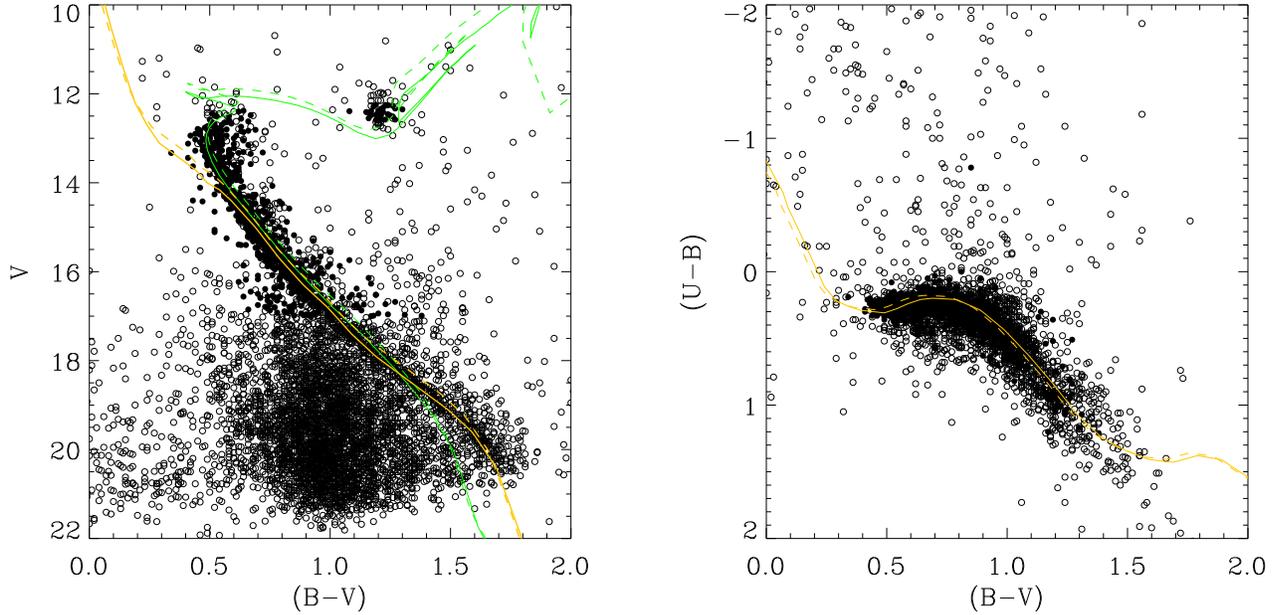}
\end{center}

\caption{Model fit results for \object{NGC~2477} (Ref. 152) where we show the rejected
  stars by our filtering method (open circles) as well as stars used
  in the fit (filled circles). The fitted isochrone and ZAMS are shown
  in thin and thick solid lines respectively as well as the values from
  \cite{PN06} with the ZAMS as well as the isochrone plotted
  with thick and thin dashed lines respectively.}

\end{figure*}

\begin{figure*}
\begin{center}
\includegraphics[scale=0.6]{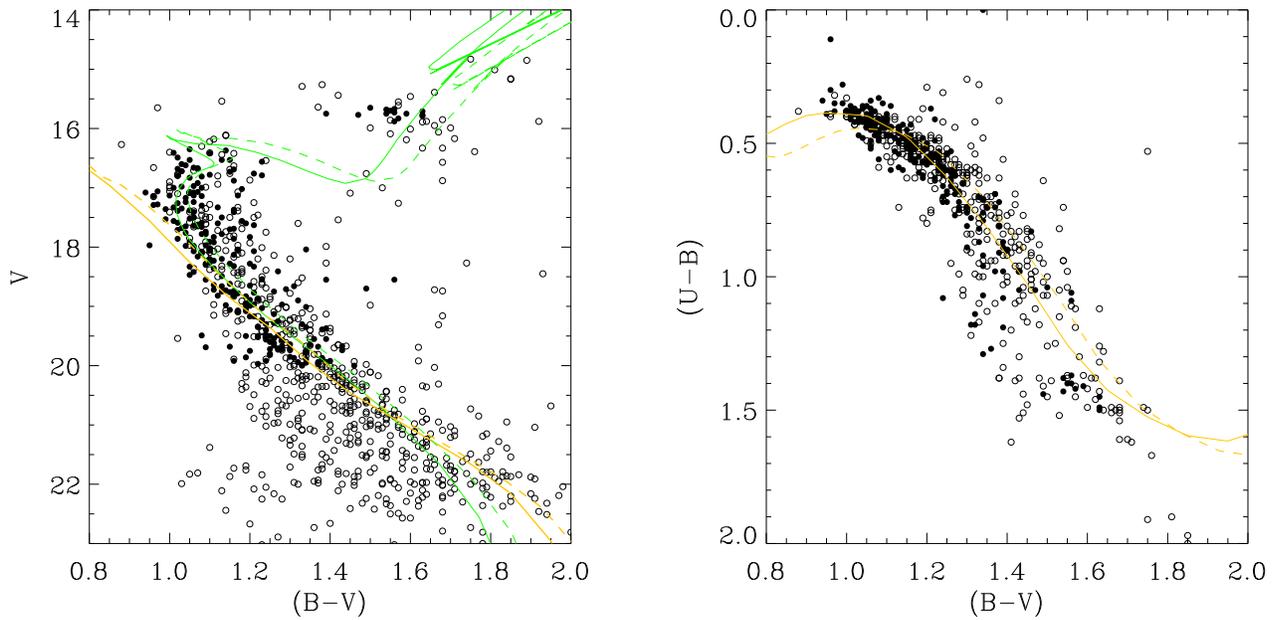}
\end{center}
\caption{Same as Fig. 7 for \object{NGC~7044} (Ref. 62).}
\end{figure*}

\begin{figure*}
\begin{center}
\includegraphics[scale=0.6]{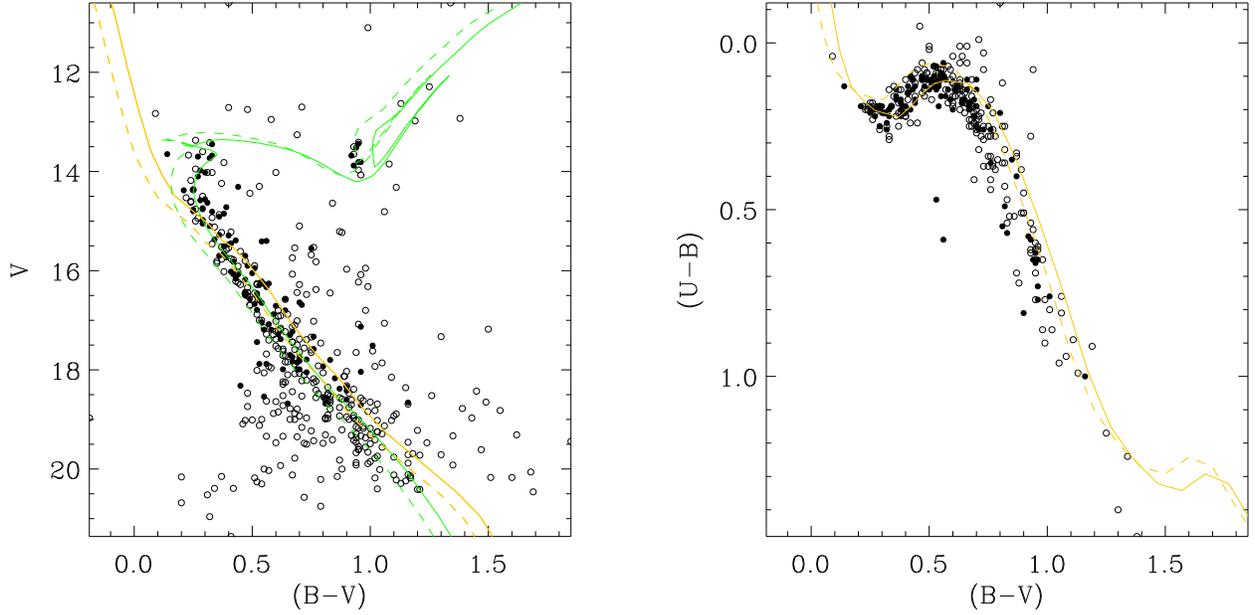}
\end{center}
\caption{Same as Fig. 7 for \object{NGC~2266} (Ref. 41).}
\end{figure*}

\begin{figure*}
\begin{center}
\includegraphics[scale=0.6]{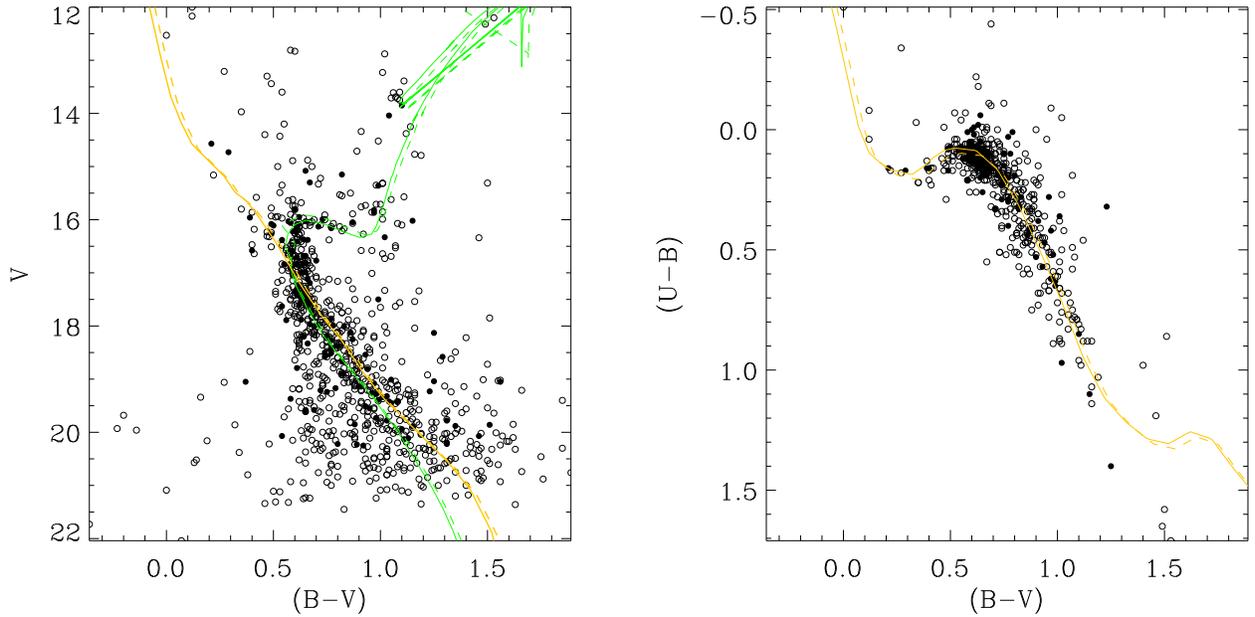}
\end{center}
\caption{Same as Fig. 7 for \object{Berkeley 32} (Ref. 40).}
\end{figure*}

\begin{figure*}
\begin{center}
\includegraphics[scale=0.6]{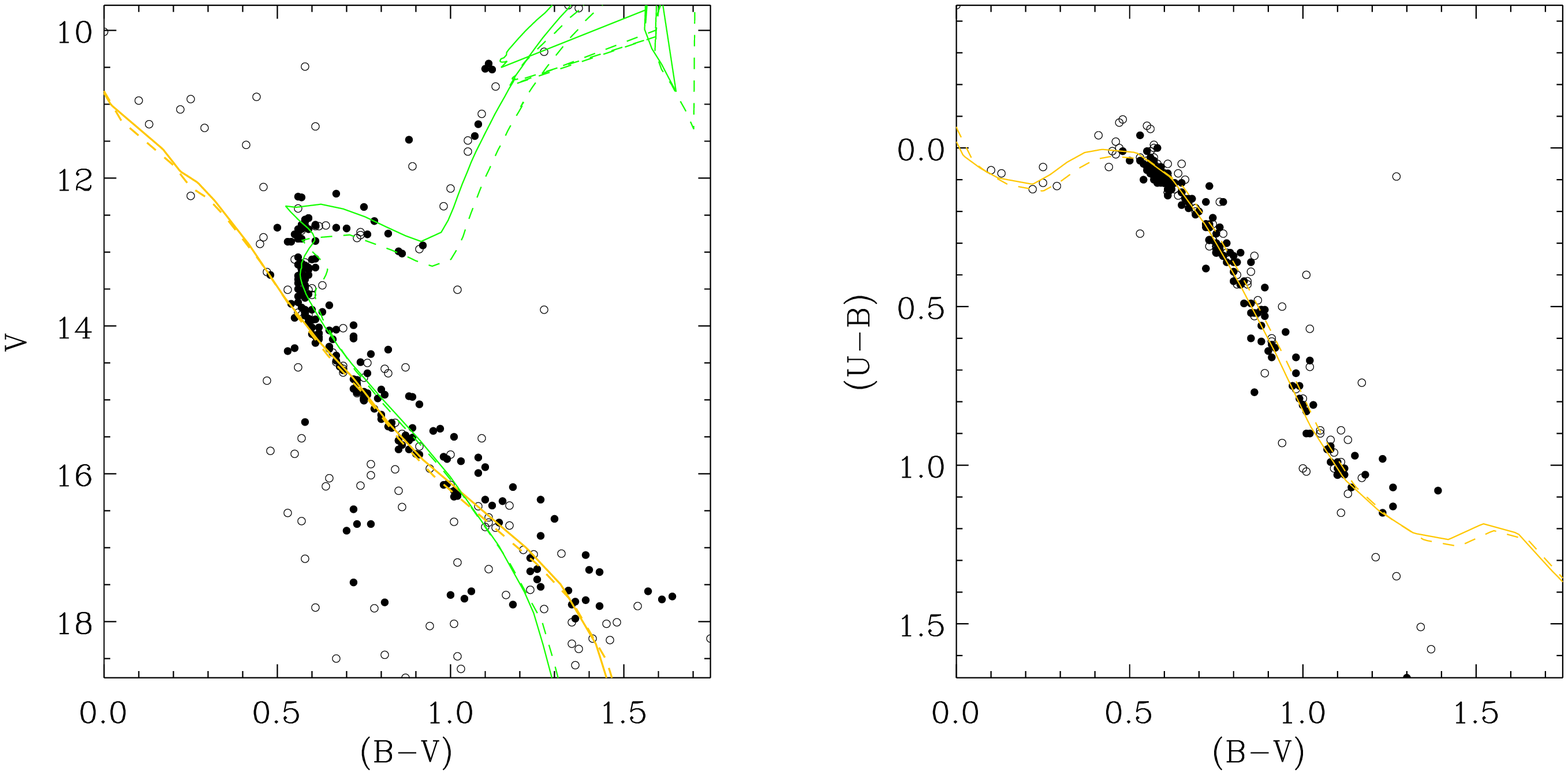}
\end{center}
\caption{Same as Fig. 7 for \object{NGC~2682} (Ref. 335).}
\end{figure*}

\begin{figure*}
\begin{center}
\includegraphics[scale=0.6]{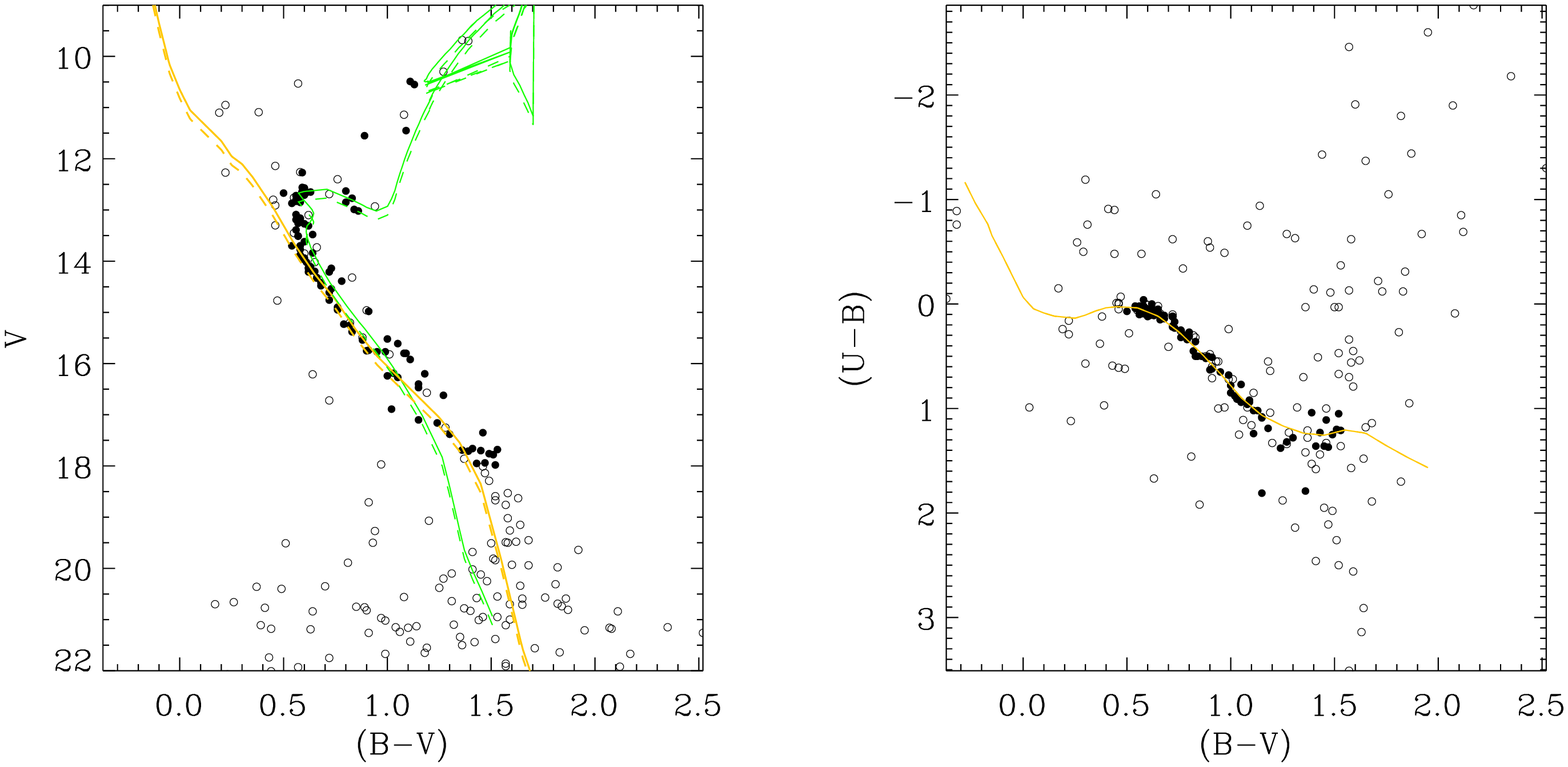}
\end{center}
\caption{Same as Fig. 7 for \object{NGC~2682} (Ref. 54).}
\end{figure*}

\begin{figure*}
\begin{center}
\includegraphics[scale=0.6]{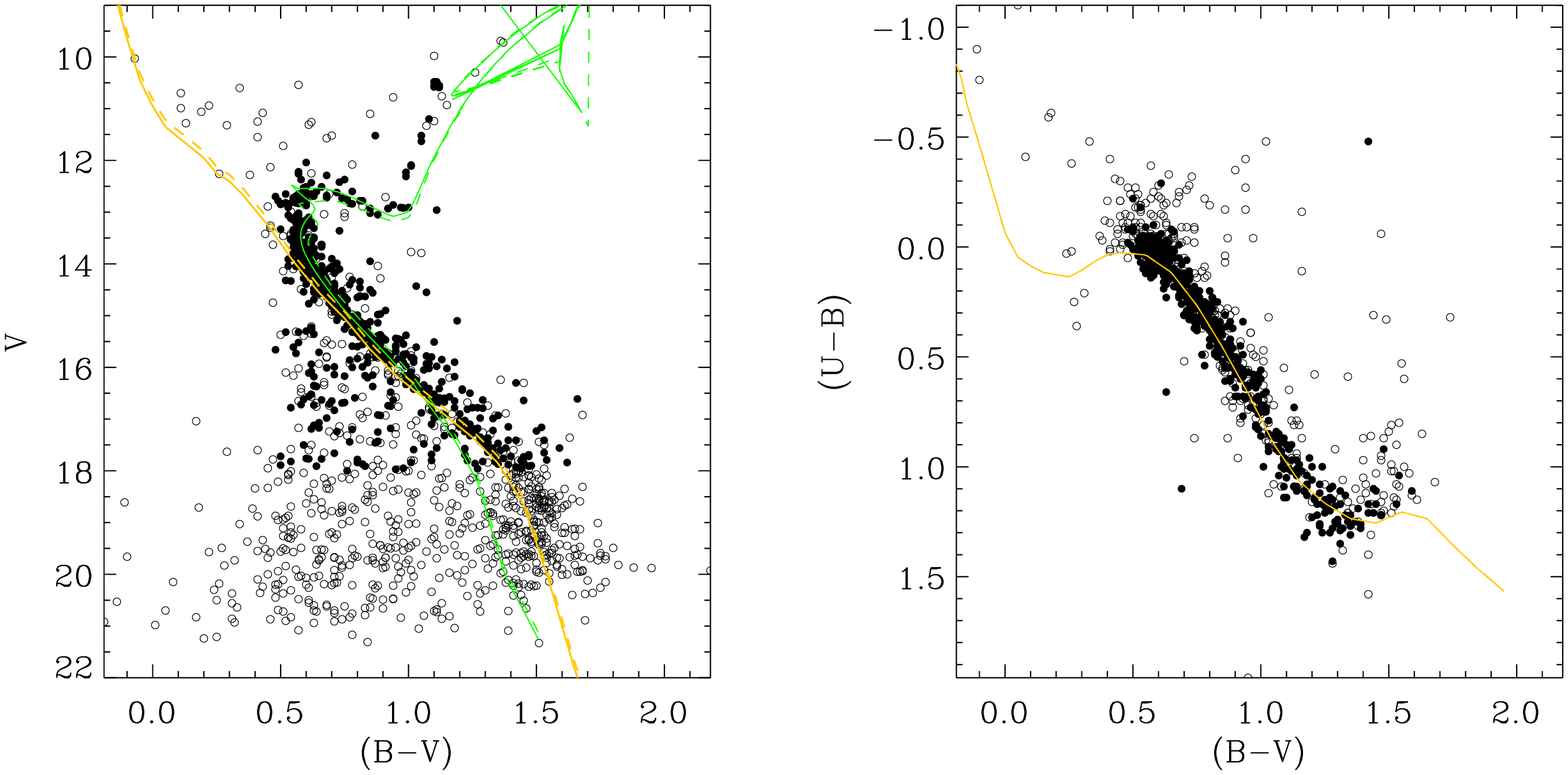}
\end{center}
\caption{Same as Fig. 7 for \object{NGC~2682} (Ref. 54).}
\end{figure*}

\begin{figure*}
\begin{center}
\includegraphics[scale=0.6]{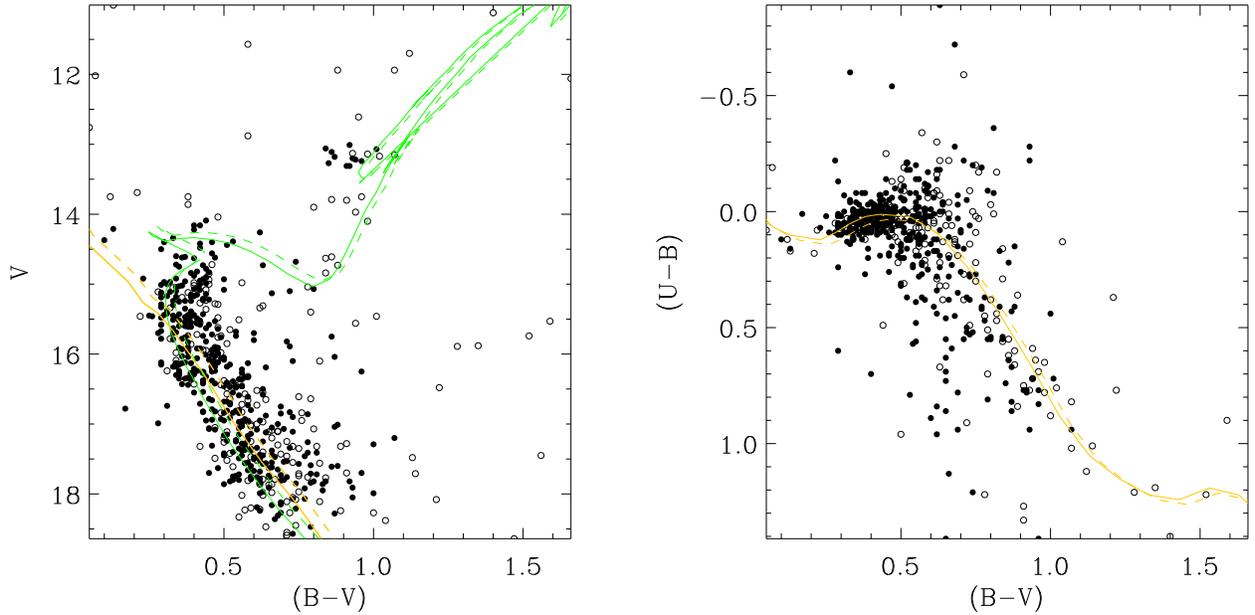}
\end{center}
\caption{Same as Fig. 7 for \object{NGC~2506} (Ref. 284).}
\end{figure*}

\begin{figure*}
\begin{center}
\includegraphics[scale=0.6]{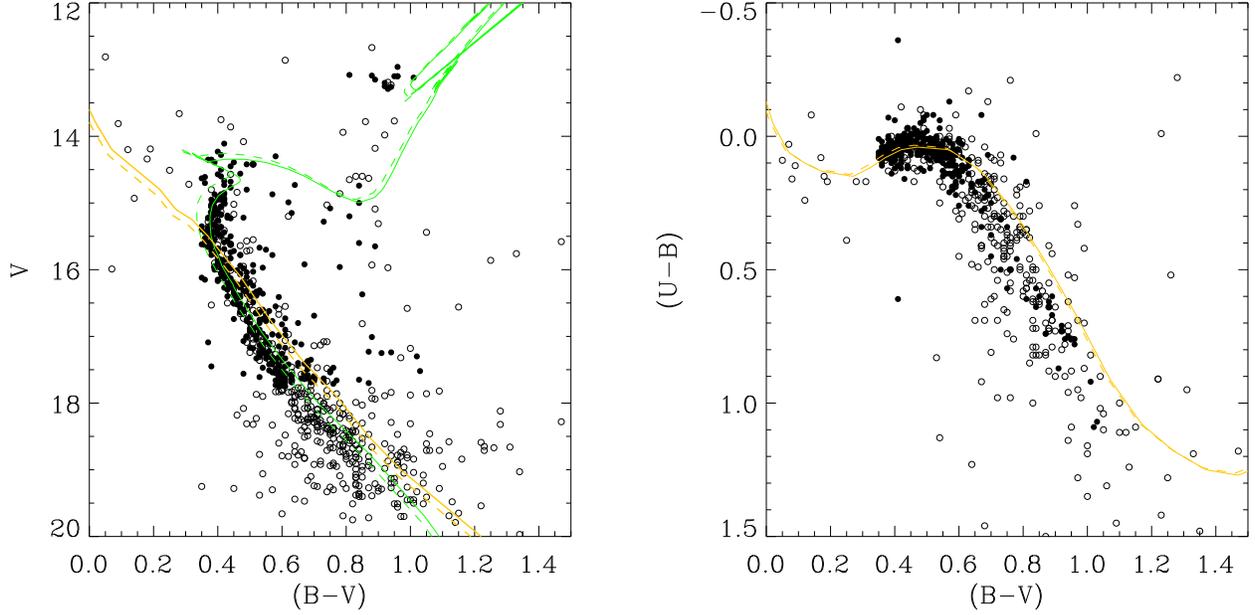}
\end{center}
\caption{Same as Fig. 7 for \object{NGC~2506} (Ref. 163).}
\end{figure*}

\begin{figure*}
\begin{center}
\includegraphics[scale=0.6]{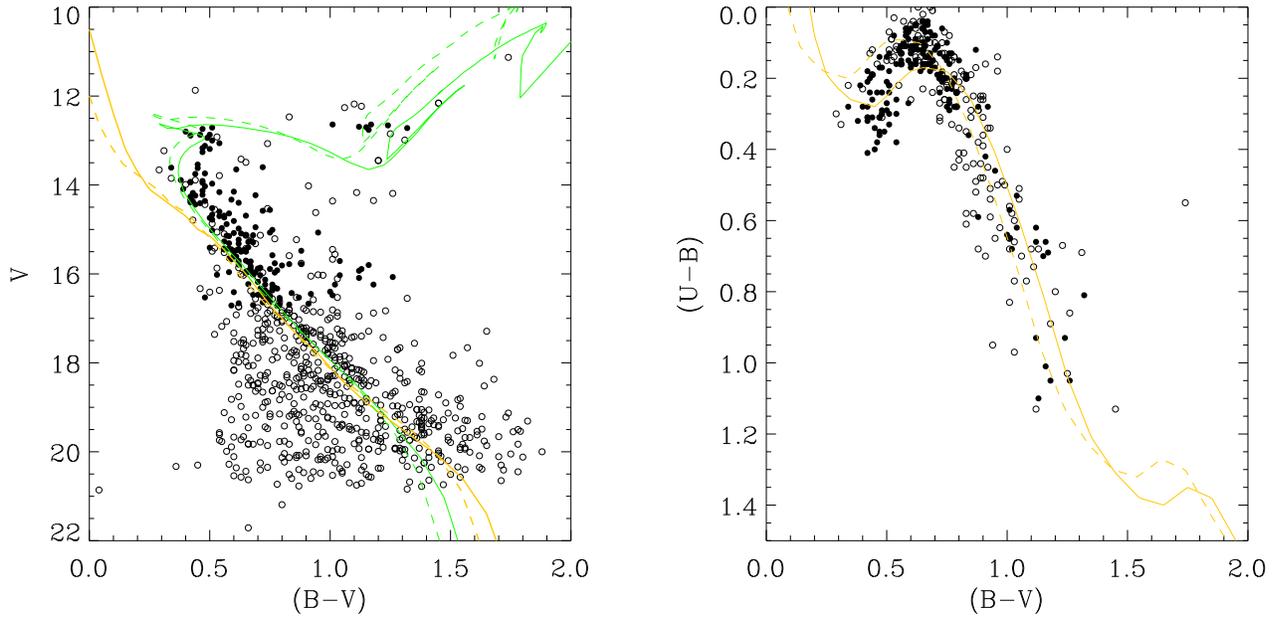}
\end{center}
\caption{Same as Fig. 7 for \object{NGC~2355} (Ref. 217).}
\end{figure*}
\begin{figure*}
\includegraphics[scale=0.6]{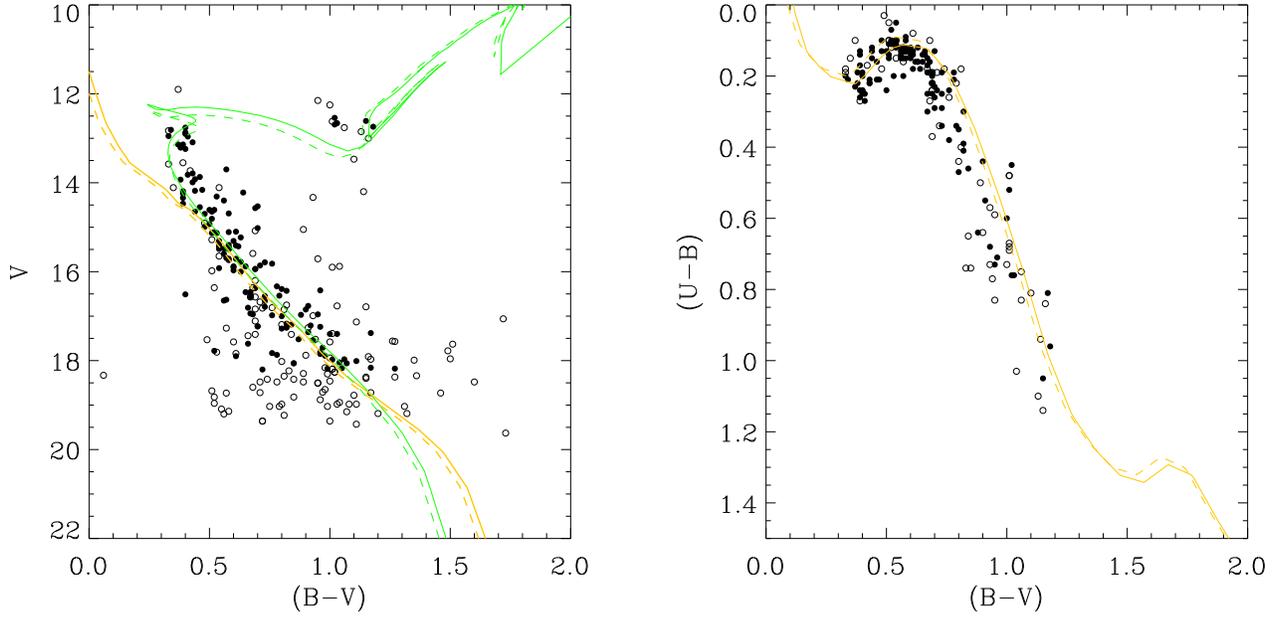}
\caption{Same as Fig. 7 for \object{NGC~2355} (Ref. 44).}
\end{figure*}

\begin{figure*}
\begin{center}
\includegraphics[scale=0.6]{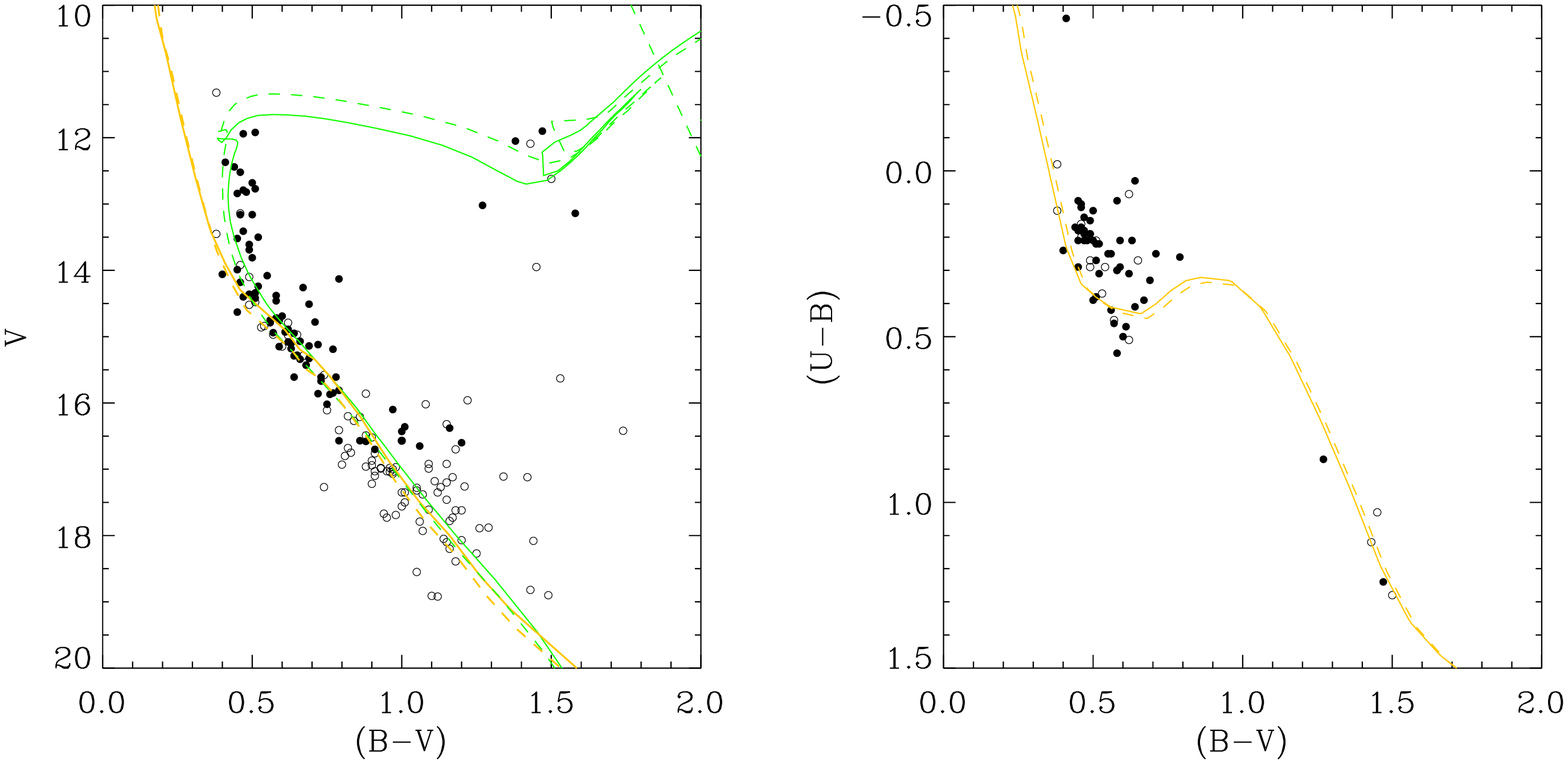}
\end{center}
\caption{Same as Fig. 7 for \object{Melotte 105} (Ref. 289).}
\end{figure*}

\begin{figure*}
\begin{center}
\includegraphics[scale=0.6]{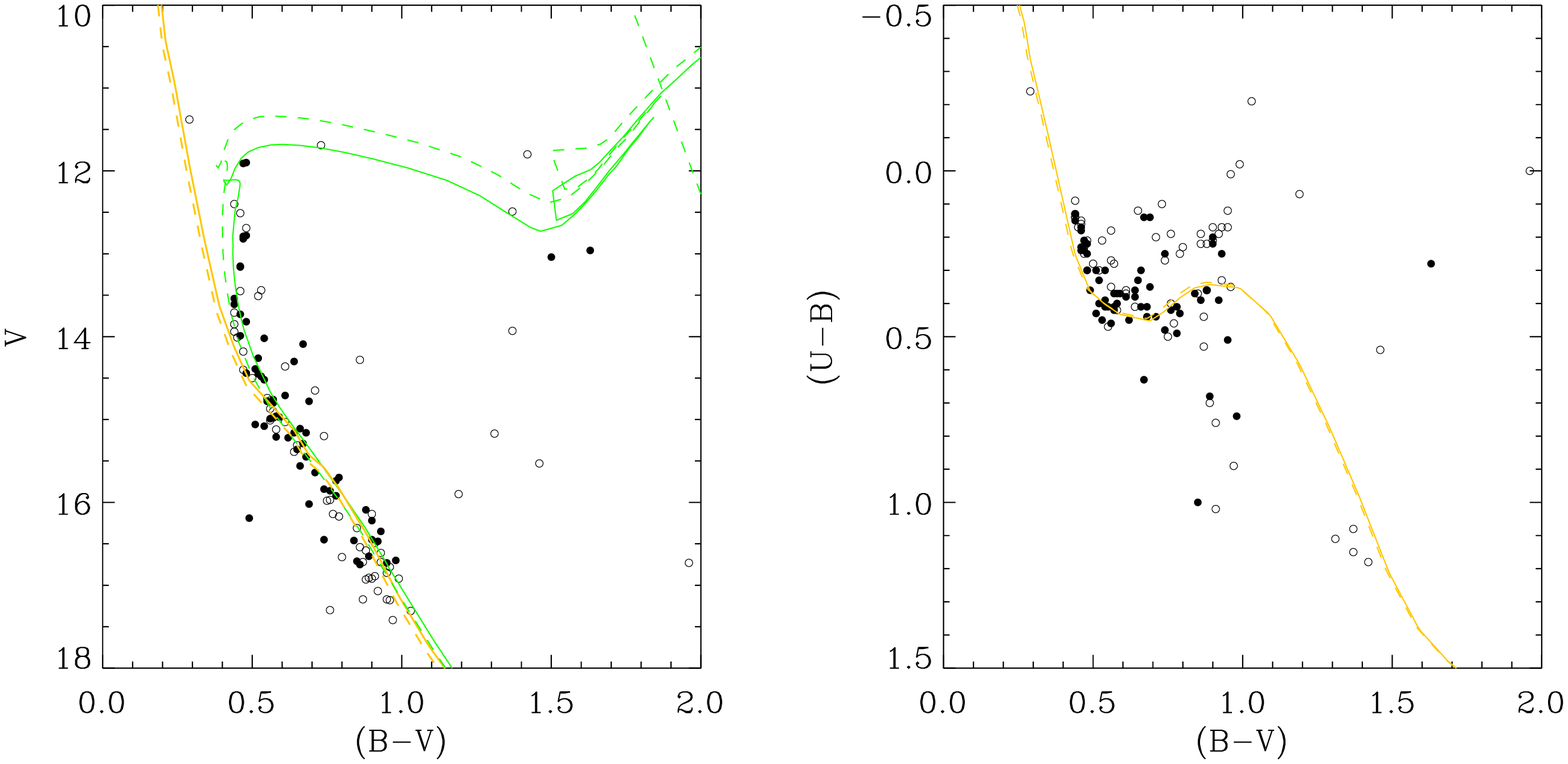}
\end{center}
\caption{Same as Fig. 7 for \object{Melotte 105} (Ref. 32).}
\end{figure*}

\begin{figure*}
\begin{center}
\includegraphics[scale=0.6]{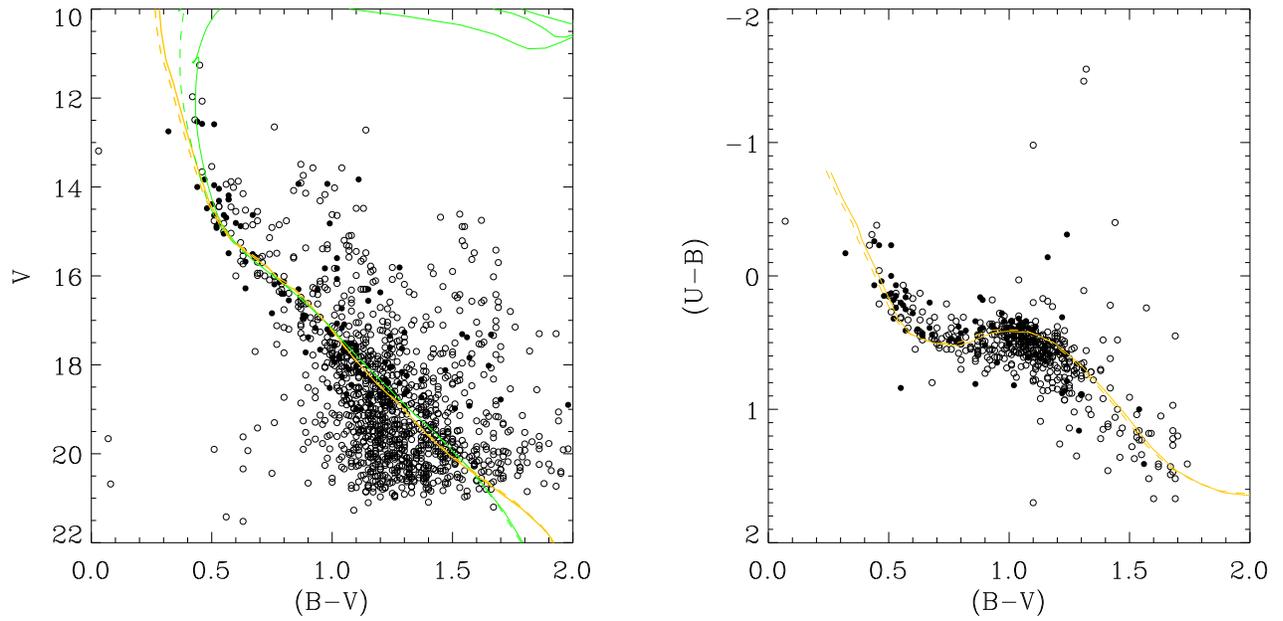}
\end{center}
\caption{Same as Fig. 7 for \object{Trumpler 1} (Ref. 320).}
\end{figure*}

\begin{figure*}
\begin{center}
\includegraphics[scale=0.6]{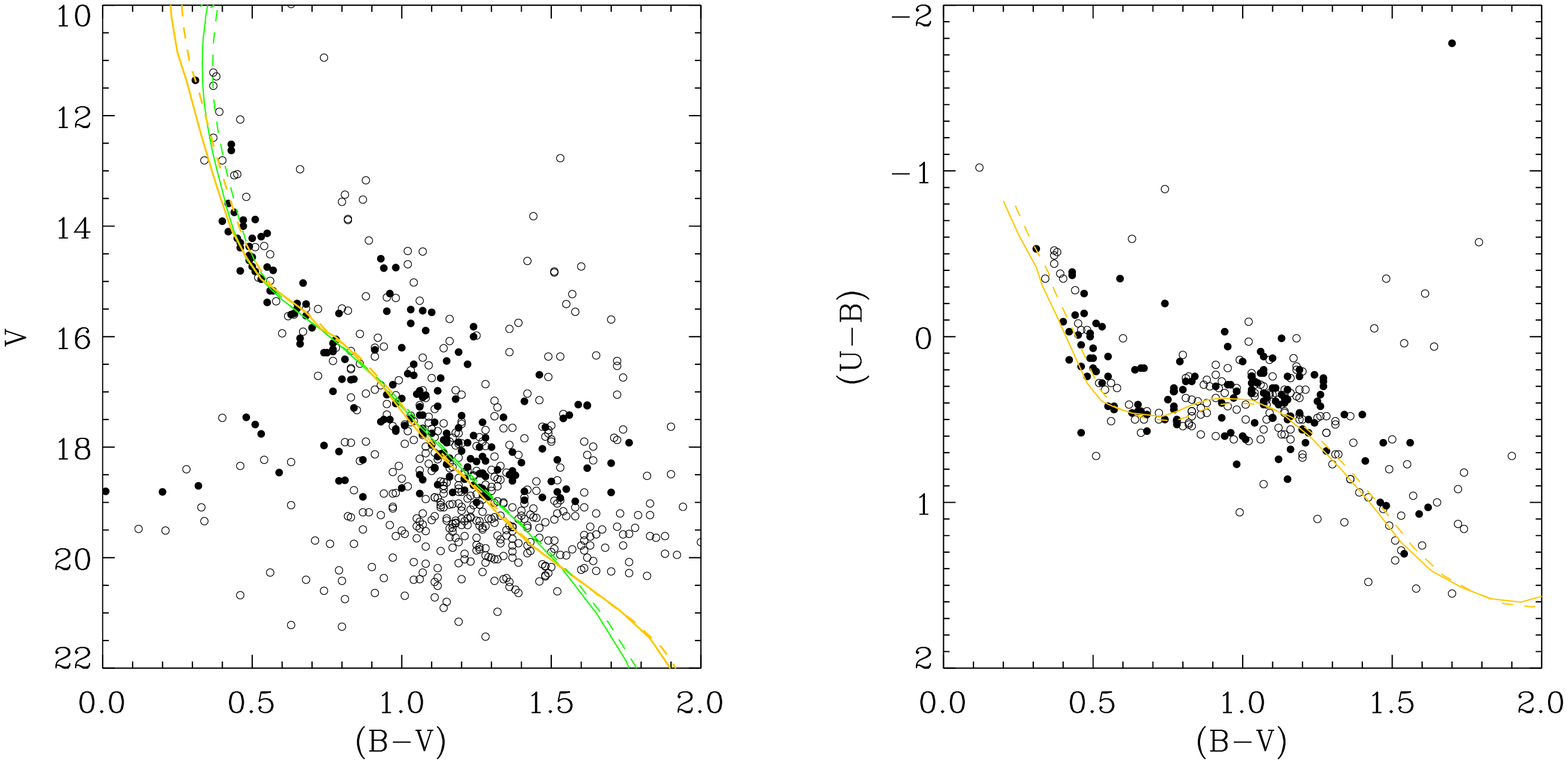}
\end{center}
\caption{Same as Fig. 7 for \object{Trumpler 1} (Ref. 86).}
\end{figure*}

\section{Conclusions}

As pointed out by \cite{PN06}, studying phenomena using open cluster
physical parameters, is highly dependent on their precision. The
authors also show that major discrepancies still exist even in well
studied clusters. In our work we provide a new technique for the
determination of open cluster physical parameters that is not
dependent on the user and is reproducible within the statistical
uncertainties given well defined conditions.

Our method, based on the Cross-Entropy optimization algorithm, was
tailored to the fitting of theoretical isochrones as the ones of
\cite{Girardi2000, Marigo2008} used in this work. The procedure is
simple and allows for the use of any tabulated theoretical isochrones
and thus provides also an unbiased means of comparing fits using
different theoretical models given the same constraints and fitting
procedure.

In this work we have concentrated in the validation of the method
limiting ourselves to fitting synthetic clusters and well studied open
clusters with the tabulated isochrones of Padova. The results show
that the method is capable of recovering the original parameters with
good accuracy even in cases where we included considerable non-uniform
field contamination demonstrating its robustness. The validation using
these synthetic clusters is by no means complete, however, we explored
the most typical situations present in open cluster data.

The results using the observed data available in the literature show
that the parameters determined through our technique are consistent
with the results obtained by other authors and especially those given
by \cite{PN06} as shown in Table 3. In all cases where we encountered
significant discrepancies these could be explained by data quality,
level of contamination, which stars were selected by our filtering
technique or some combination of these. In any case, the fact that all
the steps related to our fitting procedure are quantifiable, allows us
to perform objective comparisons of different parameters for a given
data set, removing the subjectivity of which stars are selected.

The filtering and weighting technique defined by a precise set of
conditions is central to our method. As shown in the results for
Trumpler 1, in cases where the cluster has low sampling, i.e. low
number of stars, it becomes difficult to determine the weight for
cluster stars. It is likely that better statistical tools may improve
the efficiency of this step. In general, the filtering technique
performed well in eliminating most of the contamination and assigning
weights in the observed clusters, especially in the well sampled
cases.

The final results show that there is good agreement in general with
the results adopted as standards in the literature, but also indicates
that some issues still remain unresolved. Perhaps the most important
of these issues is related to the metallicity. As mentioned before, we
kept the metallicity values fixed in our fits and attempted to use the
ones provided by the observers, except in cases where we believed more
reliable values were available. Many of the clusters show results that
could be clearly improved by changing the metallicity, as for example
the case of \object{NGC 7044}. Given the considerations above, it is also
important to point out that the values we used for comparison taken
from the proposed standard list of \cite{PN06} do not take
metallicities into account. The different author determinations for
the parameters that were averaged possibly were derived using
different metallicities.

Another aspect that plays a major role in the final results is the
binary fraction. This is a characteristic of the clusters that is not
easily accounted for even in visual fits and so it is difficult to
evaluate how much this is relevant in each individual case. We have
accounted for this effect by assuming a 100\% binary fraction and
drawing companions from the same IMF used to generate the fitting
points. While this is clearly not the correct binary fraction for all
clusters, the effect of adopting that value is relatively small, as
shown in Fig. 5, due to the fact that effectively only binary systems
with similar mass will show a significant difference in magnitude on
the CMD.

Given all the consideration above, we show that our method is reliable
and robust and although the results presented in this work are
consistent with literature values all clusters we believe that there
is room for improvement in the accepted parameter values for other
clusters. The possibility of re-evaluating previous results with more
quantifiable means is important as it removes the subjectivity
inherent in most open cluster studies up to today.

\section{Acknowledgments} 

We thank Professor Jacques L\'epine for motivation and for some
helpful discussions. We also thank the referee Dr. Bailer-Jones for
extremely useful comments that greatly improved the paper. W. S. Dias
thanks to CNPq (grant number 302762/2007-8), CAPES (project
CAPES-GRICES process 040/2008) and FAPEMIG (process number
APQ-00090-08). T. C. Caetano thanks to CAPES. Extensive use has been
made of the WEBDA databases.

\bibliographystyle{aa} 
\bibliography{myrefs.bib} 

\end{document}